\documentclass[draft]{agujournal2019}
\usepackage{url} %this package should fix any errors with URLs in refs.
\usepackage[utf8]{inputenc}
\usepackage[inline]{trackchanges} 
\usepackage{natbib}
\usepackage{soul}
\usepackage{graphicx}
\usepackage{multirow}
\usepackage{helvet}
\usepackage{amsmath}
\usepackage{color}
\usepackage{colortbl}
\usepackage{setspace}
\usepackage{outlines}
\usepackage{cancel}
\usepackage{todonotes}
\draftfalse
\graphicspath{{Figures_low/}}
\journalname{JGR: Space Physics}
\addeditor{sb}
\addeditor{gg}
\addeditor{sh}
\newcolumntype{a}{>{\columncolor{green!15}}c}
\newcolumntype{x}{>{\columncolor{blue!15}}c}
\newcolumntype{s}{>{\columncolor{red!15}}c}
\newcolumntype{y}{>{\columncolor{yellow!15}}c}
\begin{document}
\begin{singlespace}
	\title{The Sun-Earth-Moon Connection:\\ I--3D Global Kinetic Simulation}
	\authors{Suleiman M Baraka \affil{1,6}, Sona Hosseini \affil{2},  Guillaume Gronoff\affil{3,7}, L. Ben$-$Jaffel\affil{4}, Robert Rankin\affil{5}}
	\affiliation{1}{National Institute of Aerospace, 100 Exploration Way, Hampton, VA 23666}
	\affiliation{2}{Jet Propulsion Laboratory, California Institute of Technology, M/S 183–401, 4800 Oak Grove Drive, Pasadena, CA 91109, USA}
	\affiliation{3}{Chemistry and Dynamics Branch, Science Directorate, NASA Langley Research Center, Hampton, Virginia, USA}
\affiliation{4}{Institut d’Astrophysique de Paris, UMR7095, Sorbonne Université, Paris, France}
	\affiliation{5}{Department of Physics, University of Alberta, Edmonton, Alberta, Canada,}
	\affiliation{6}{Department of Physics and Astronomy, University of Calgary, Calgary, Alberta, Canada.}
	\affiliation{7}{Science Systems and Application Inc, Hampton, Va, USA}
	\correspondingauthor{Suleiman M Baraka}{suleiman.baraka@nianet.org}
	\begin{keypoints}
 \item A Particle In Cell 3D Simulation of the Sun-Earth-moon system was performed, allowing to derive the magnetopause shape and size
\item The simulation allowed the characterization of the backstreaming ions from the solar wind 
\item The hydrogen and oxygen ion escaping from the Earth's ionosphere have been tracked to the Moon's environment
	\end{keypoints}
\end{singlespace}
\justify
\begin{abstract}
\textbf{Context:}
The complex interplay between the Solar Wind and the lunar surface serves as a quintessential example of space weathering. However, uncertainties persist regarding the influence of plasma originating from Earth's ionosphere, necessitating a comprehensive understanding of its quantitative impact. Hitherto, the dearth of reliable models has impeded accurate computation of ion flux from Earth to the Moon under varying solar wind conditions.
\textbf{Aim:} The objective of this study is to adapt a kinetic model for the challenging conditions of having both the Earth and the Moon in a single simulation box.
\textbf{Methods:}  $\mathrm{IAPIC}$, the Particle-In-Cell Electromagnetic Relativistic Global Model was modified to handle the Sun-Earth-Moon system. It employs kinetic simulation techniques that have proven invaluable tools for exploring the intricate dynamics of physical systems across various scales while minimizing the loss of crucial physics information such as backscattering. 
\textbf{Results:} The modeling allowed to derive the shape and size of the Earth's magnetosphere and allowed tracking the O$^+$ and H$^+$ ions escaping from the ionosphere to the Moon: $\mathrm{O^+}$ tends to escape towards the dayside magnetopause, while $\mathrm{H^+}$ travels deeper into the magnetotail, extending up to the Lunar surface. In addition, plasma temperature anisotropy and backstreaming ions were simulated, allowing for future comparison with the experiment.
\textbf{Conclusion:} This study shows how a kinetic model can successfully be applied to study the transport of ions in the Earth-Moon environment. A second paper will detail the effect on the Lunar environment and the impact on the Lunar water.
\end{abstract}
\clearpage
\section{Introduction}\label{intro}
The Sun-Earth-Moon system is at the center of attention now that multiple space missions are planned for the next step of humans on the lunar surface. This endeavor brings new light to the question of the water at the surface of the Moon and its origins, as well as of the effect of plasma on the surface of the satellite and its implications for space missions \citep{Dandouras2023}. It has indeed been shown that the hydrogen coming from the solar wind impacts the lunar rocks to create water \citep{Hurley2000}. Moreover, the lunar regolith is sensitive to electric fields, which may be affected by the solar wind \citep{Zakharov2020}. Since the Moon periodically enters the Earth's magnetotail, it has been suggested that its plasma environment is greatly changed: does it reduce the risk of surface charging? Does it change the amount of water created by space weathering, for example by bringing O$^+$ \citep{Terada2017}?
To answer these questions, it is necessary to adapt a plasma model to the Sun-Earth-Moon system and to study in detail the effects near the Lunar surface. We will do that through a series of papers, each focused on a specific aspect of the model results.
Paper I investigates the interactions between the solar wind, magnetosheath, and magnetosphere, as well as the coupling between the magnetosphere and ionosphere. It reports on the transport of $\mathrm{H^+}$ and $\mathrm{O^+}$ particles originating from the upper ionosphere of the Earth. These particles are tracked both in the day side and night side of the magnetosphere, extending to the lunar surface within the magnetotail.
Paper II details the  interaction between the solar wind and the Moon within the magnetotail. The lunar surface's charging, its wake dynamics, and the characterization of backstreaming ions and temperature anisotropy are reported in detail. 
In the present paper, we, therefore, concentrate on the magnetopause shape and on the transport of ions coming from both the solar wind and the Earth's ionosphere,  in the case when the Sun, the Earth, and the Moon are aligned.
Several parameters have to be studied in detail to understand how the plasma affects the lunar surface. 
First, the solar wind is the primary driver of energy transfer to the magnetosphere and the cause of almost all its dynamics. It is a plasma originating from the Sun, which is characterized by its density, speed, and its magnetic field, also called Interplanetary Magnetic Field (IMF).
The energy transfer relates to the IMF and the large-scale reconnections at the dayside magnetosphere \citep{Tenfjord2013}. Magnetic Reconnection is a kinetic process through which the magnetic field energies are converted to particle kinetic energies, which results in the effective exchange of mass momentum and energy between different magnetized plasma regions \citep{Paschmann2008}. 
Second, the magnetopause (MP), i.e. the boundary between the solar wind and the plasma within the magnetosphere, determines the specific area where the incoming solar wind is deposited  in the inner magnetosphere \citep{Zhang2022}. At the subsolar point, the location of the magnetopause is determined by the balance between the planetary magnetic field pressure and the dynamic pressure of the solar wind, as predicted by the classical fluid description of the solar wind stagnation flow. Plasma boundary layers form on either side of the magnetopause, with the magnetosheath boundary layer (MSBL) on the sunward side and the low-latitude boundary layer (LLBL) on the magnetosphere side. These boundary layers play an essential role in plasma exchange across the magnetopause \citep{Pi2018}.  During the northward IMF, the global dynamic of the magnetosheath and the magnetosphere changes due to the high latitude magnetic reconnections at the Earth's magnetopause. The exchange of solar wind particles reflects the magnetosheath-magnetosphere coupling\citep{Fuselier2012}. While on the other hand, at the subsolar magnetopause, the reconnection between the magnetosheath and the magnetosphere occurs during southward IMF \citep{Trattner2007,Fuselier2011}. More details about the spatial and temporal scales of reconnection locations and particle accelerations are reported in  \citet{Paschmann2008} and the references therein. The IMF orientations directly impact the magnetopause structure, and the influence of southward IMF is reported in several studies \cite[e.g.]{Yu2009,Heikkila2011,Tan2011,Suvorova2015,Berchem2016} and for the northward IMF  \citep{Onsager2001,Luo2013,Bobra2004,Wang2018b,Sorathia2019,Wang2018,Fear2021,Wang2021}. \citet{Wiltberger2017} reports that the high latitude currents are flowing out of the ionosphere on the Dusk side, and the low latitude currents are flowing into the ionosphere to account for the magnetosphere-ionosphere coupling. These two processes are reversed for the Dawn side of the ionosphere. The basis of the electrodynamics linkage between the magnetosphere and ionosphere is referred to as the field-aligned currents (FAC) \citep{Birkeland1908}. Finally, once we understand the transfer of energy and solar wind plasma in the magnetosphere, it is important to understand that some plasma can be injected into the magnetosphere through the escape of ions from the Earth's upper atmosphere. 
From there, we have all the pieces of the puzzle to understand how the space weather can affect the lunar surface in the case of the alignment: the solar wind affects the magnetosphere's shape and energetic, in response, it changes density and direction, which can affect its interaction with the Moon; in addition, the energy impeded into the magnetosphere affects its plasma (from the Earth's ionosphere or solar wind) which can be transported in the magnetotail and the Moon's vicinity.
Furthermore, the distribution of escaping \(\mathrm{O^+}\)   ions from the upper ionosphere exhibits a distinct morphological asymmetry between the Dusk and Dawn sectors of the equatorial plane. This  asymmetry  is particularly  prominent during periods of high solar wind dynamic pressure, especially when IMF is oriented northward as reported in paper I. This observation has been documented through statistical analysis of data collected by the Magnetospheric Multiscale (MMS) mission, as reported by \cite{Zeng2020}. The distribution of \(\mathrm{H^+}\)  and \(\mathrm{O^+}\) in the near-Earth region will be compared to Cluster observations as reported in \citep{Kronberg2015}.  The distribution of \(\mathrm{H^+}\)  and \(\mathrm{O^+}\)  density and temperature will compered with observations reported in \citep{Wing2005}

In Section 2, we will describe the IAPIC model that is used to perform these calculations as well as the conditions kept for the simulation. In Section 3, we will present the results of the model, first by highlighting the shape of the magnetosphere and then by describing the effect on the ion density, before discussing these results in light of previous observations (Section 4) and concluding.
\clearpage
	\justify \section{Initial conditions and Simulation Model: IAPIC}\label{code}
 The Particle-In-Cell Electromagnetic Relativistic Global Model, $\mathrm{IAPIC}$, \citep{Baraka2021} was used for this study. It incorporates boundary conditions and employs charge conservation principles, as discussed in \citet{Lindman1975, Villasenor1992}.
 
{The present simulation models the full Moon at the Sun-Earth Line's far end. This is, to our knowledge, the first time the Earth and the Moon have been included in a single simulation box that preserves physical parameters. The  box's dimensions are chosen to be \(495 \times 225 \times 225 \mathrm{\Delta x} =99 \times 41 \times 41 \mathrm{R_E}\) with the spatial resolution of our cell size \(\mathrm{\Delta x}\)=0.2\(\mathrm{R_E}\). These parameters were chosen to encompass both the Moon and the Earth within the same simulation and to capture the entire designated regions of the global macrostructure of the Earth's magnetosphere, ensuring sufficient distance within the magnetotail. The Earth is positioned at \(30\mathrm{R_E}\) away from the simulation box boundary sunward, and the Moon is located \(60\mathrm{R_E}\) away from the Earth along the simulation axis. The simulation box contains $1.3\times10^{8}$ solar wind pair particles (\(i.e.\ n_i=n_e=1.3\times10^{8}\)) and \(1.\times 10^{7}\) hydrogen \(\mathrm{H^{+}}\) and oxygen \(\mathrm{O^{+}}\) ions of ionospheric origin, with an ion-to-electron mass ratio of 100 for the solar wind. This number of particles is chosen based on a scaling factor of the number of cells in the simulation box which is 5 pair of particles per cell. To simulate the Magnetosphere-ionosphere coupling, a spherical symmetric ionospheric model (International reference ionosphere $\mathtt{IRI-2007}$) was incorporated into the 3D PIC EM Global code, as described in \citet{Bilitza2008, Bilitza2011}. To explore the dynamics and time-dependent behavior of the 3D magnetosphere, the $\mathtt{IRI-2007}$ model was integrated with the magnetospheric code to investigate the impact of the supply of thermal ions (i.e., \(\mathrm{H^{+}}\) and \(\mathrm{O^{+}}\) ions) from the upper ionosphere on the magnetosphere's content and behavior. By accommodating these celestial bodies within a single box, we can effectively study their interactions and dynamics comprehensively. It is important to note because we utilize the same simulation box for the moon and Earth, due to the difference in length scale between the Sun-Earth and Earth-Moon, visualizing the moon accurately is challenging; the moon's radius \(\mathrm{R_L}\) being 0.2727 \(\mathrm{R_E}\). To address this issue, we have  scaled  up the \(\mathrm{R_L}\) to \(\approx\) 5 times its original size, which corresponds to \(\approx\) 1.0\(\mathrm{R_E}\) to allow for a clearer visualization of the moon within the simulation box without causing global-scale impacts.}\par

{For this study, the initial conditions were set based on well-established PIC code simulations physical parameters \citep[e.g.,][]{Cai2006,Nishikawa2003,Ben-Jaffel2014}. We use a supersonic solar wind speed that is equivalent to a steady flow speed of 500 km/sec (code value 0.25) and a normal condition solar wind density of 5n/$\mathrm{\Delta}$, supplied by a northward interplanetary magnetic field (IMF) of 2.2 nT (code value 0.2), in order to achieve the expected magnetopause length. Solar wind particles that reach the Moon were allowed to accumulate (rather than being removed, as in \citep[e.g.,][]{Poppe2014} to account for charge separation and potential differences on the lunar surface. The initial plasma parameters are calculated at $\mathrm{\Delta t}=0$ and their evolution is followed  at $t=3700\mathrm{\Delta t}$ for the current simulation run which are summarized in Table \ref{table1}.}\par 

\begin{table}
	\caption{The solar wind parameters used at the initial state and after $t=3700\Delta t$ in the undisturbed solar wind for both ions and electrons with a northern IMF are presented}\label{table1}
	\resizebox{1.\textwidth}{!}{\begin{tabular}{lll}
			\hline  \\[0.3pt]
		\underline{\underline{Parameters}}/Time step &$  t=0 \Delta t  $& $ t=3700 \Delta t  $\\ \hline  \\[.53pt]
			Ion plasma-frequency, $\omega_{ci}$    &   0.002  &   0.005\\
			Electron plasma-frequency, $\omega_{ce}$    &   0.250  &   0.480\\
			Ion gyro-frequency, $\omega_{pi}$    &   0.035  &   0.031\\
			Electron gyro-frequency, $\omega_{pe}$    &   0.349  &   0.314\\
			Ion gyro-radius, $\rho_{i}$    &  100.00  &   16.35\\
			Electron gyro-radius, $\rho_{e}$    &    1.00  &    0.19\\
			Ion inertial length, $d_{i}$    &   14.31  &   15.89\\
			Electron inertial length, $d_{e}$    &    1.43  &    1.59\\
			Ion beta, $\beta_{i}$    &    0.16  &    1.30\\
			Electron beta, $\beta_{e}$    &    0.16  &   31.71\\
			Ion Debye length, $\lambda_{Di}$    &    0.29  &    0.43\\
			Electron Debye length, $\lambda_{De}$    &    0.29  &    0.68\\
			Alfven velocity, $v_{A}$    &    0.04  &    0.03\\
			Sound speed, $c_{S}$    &    0.01  &    0.11\\
			Mach number, $M_{A}$    &    7.02  &    2.62\\
			Sonic Mach number, $M_{s}$    &   17.77  &    5.58\\
			Magnetosonic Mach number, $M_{ms}$    &    6.53  &    2.05\\
			Frequency ratio, $\omega_{p}/ \omega_{c}$   &    1.40  &    0.65\\
			\hline  \\[1pt]
			Loaded simulation box information&&\\ \hline  \\[1pt]
			\multicolumn{2}{l}{Grid Size}&$\Delta=0.2R_E$ \\
			\multicolumn{2}{l}{Time step}&$\Delta t=\Delta x/ \Delta v$ $\approx 0.64 sec$\\
			\multicolumn{2}{l}{Simulation box size}&$ (495\times 225 \times 225)\Delta $\\
			\multicolumn{2}{l}{\# of SW pair particles}&$1.3\times 10^{8}$\\
			\multicolumn{2}{l}{\# $ H^{+} $+ $ O^{+} $ ions}&$2.0\times 10^{7}$\\
			\multicolumn{2}{l}{Ion to electron mass ratio}& 100\\
			\multicolumn{2}{l}{SW ion density }& $ n_i=n_e=5/\Delta $\\\hline
	\end{tabular}}
\end{table} \clearpage
\clearpage
\section{Simulation Results}\label{results}

{The Magnetopause (MP) is a region within the magnetosphere where energy, momentum, and mass from the solar wind are transferred into the inner magnetosphere, especially when the dayside transients are considered\citep{Haaland2019}. The influence of ionospheric ions, specifically \(\mathrm{H^+}\) and \(\mathrm{O^+}\), in the magnetosphere-ionosphere coupling regime reveals diverse distributions of these species on both the day and night sides of the magnetosphere. These variations have the potential to modify the overall dynamics of Earth's magnetosphere on a global scale. Additionally, the process of the magnetosphere (MS)-Ionosphere(IS) coupling  results in tracking the \(\mathrm{H^+}\) and \(\mathrm{O^+}\) in the magnetotail, ultimately following them to the lunar surface \citep{Toledo2021,Liuzzo2021} (subsection \ref{result2}).}\par 
%%%%%%%%%%%%%%%%%%%%%%
\subsection{The Solar Wind-Magnetosphere Coupling}\label{result1}
%%%%%%%%%%%%%%%%%%%%%
The magnetopause's size was determined using the pressure balance method, as described in \citep{Baraka2021}. The IAPIC model is based on Cartesian grid coordinates. Still, to obtain a more accurate sizes estimation that accounts for planet tilt, the Cartesian 3D simulation box quantities (density, velocity vector, etc. at (x,y,z) positions) had to be transformed into a spherical 3D domain (same quantities at (i.e., r=8.25$\mathrm{R_E}$, $\theta=0^{\circ}$, $\phi=-180^{\circ}$)). However, this practical transformation does not limit the area of focus, which is the magnetosphere on the daytime side.\par
These simulations rely on scaled values derived from in-situ observational data (i.e., ARTEMIS data \citep[e.g.,][]{Angelopoulos2009}). This is primarily due to the lack of available resources to accommodate the immense distances and real plasma parameters associated with space, i.e., $v_{sw}$=500 km/sec($\equiv 0.25$ in the model), a solar wind density of $n_{i,e}=5~ n/cc$, and a northward IMF of ~2.2 nT ($\equiv 0.2$ code value).\par
\begin{figure}[ht]
	\resizebox{1.\textwidth}{!}{\begin{tabular}{cc}
			
			\includegraphics[width=0.9\linewidth]{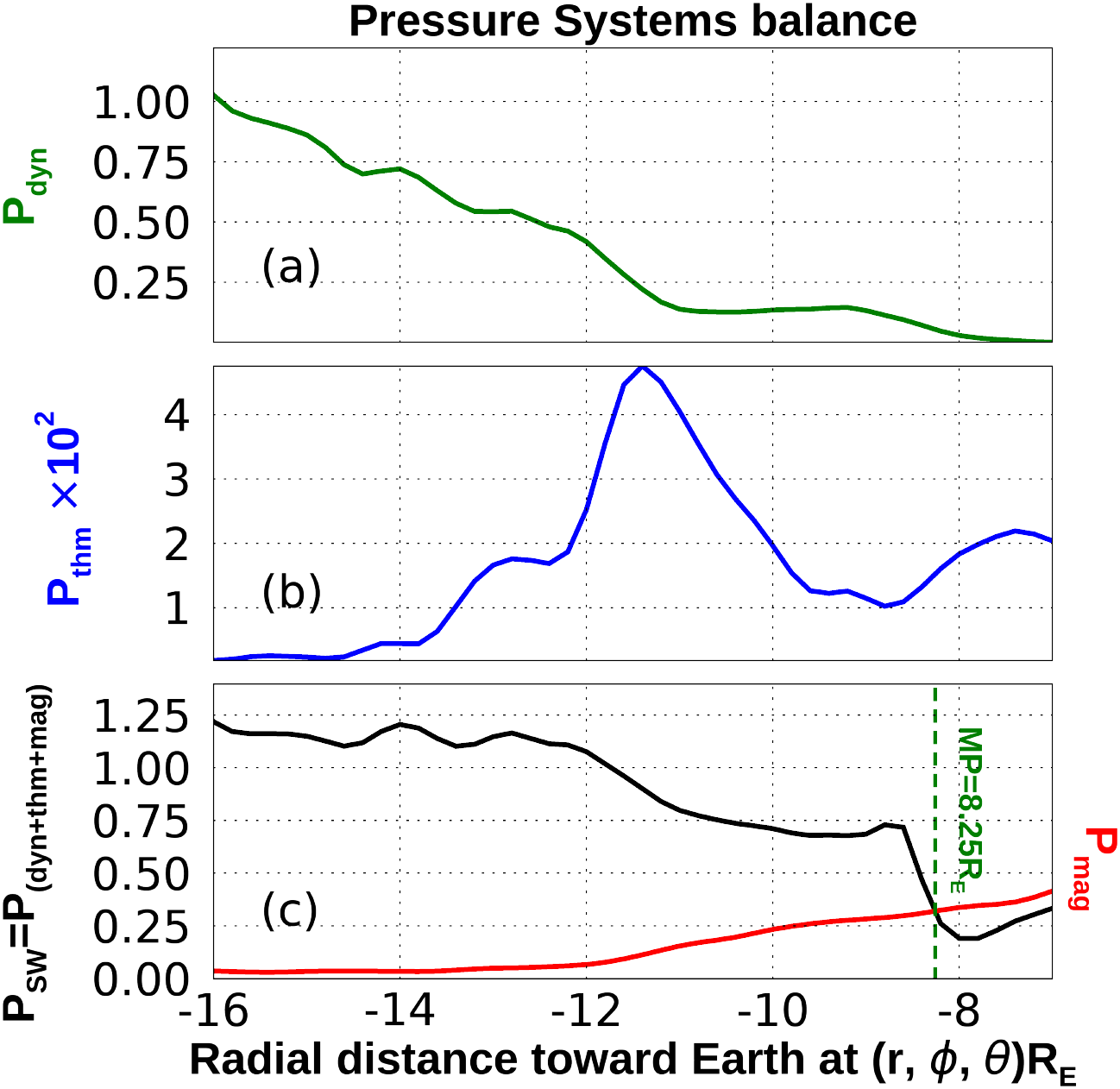}	& \includegraphics[width=0.9\linewidth]{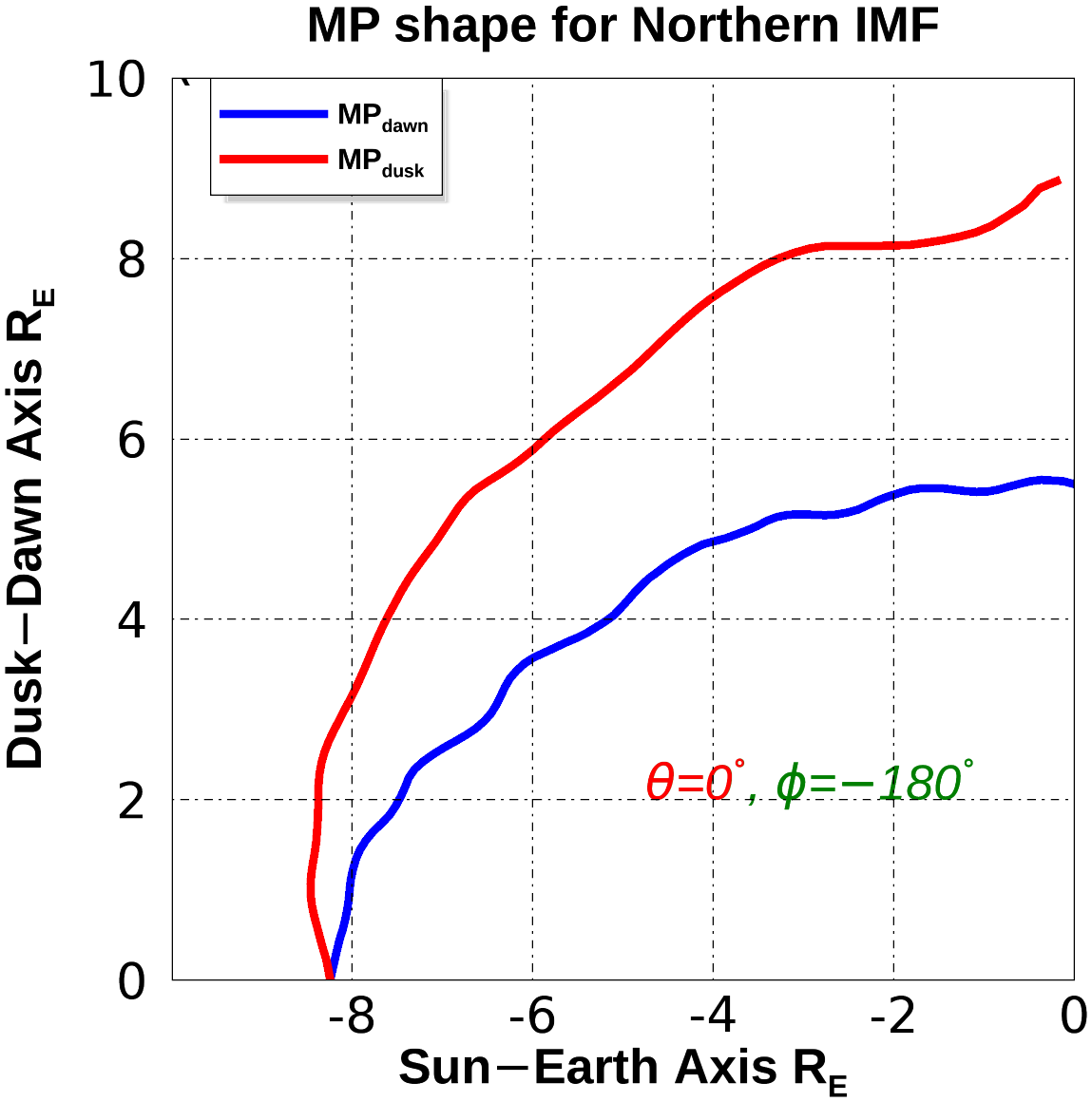}\\
			A&B\\ 
	\end{tabular}}
	\caption{\fontfamily{cmtt}\selectfont The dimensions and form of the Earth's Magnetopause (MP) are determined through a series of measurements. Panel-A demonstrates this process through an (a) dynamic pressure profile, (b) thermal pressure profile, and (c) total pressure profile (thermal, dynamic, and magnetic) in the solar wind, which is then balanced with the magnetic pressure of the dipole. The size of the MP is determined to be 8.25$R_E$. In Panel B, the shape of the MP is depicted in a single plane to consider Dusk-Dawn asymmetry.}
	\label{mpderivation}
\end{figure}

{By examining the point at which the dipole pressure intersects with the total pressure of the solar wind, one can determine the size of the MP. The MP size derived by the pressure balance method equals to 8.25\(\mathrm{R_E}\). In Figure \ref{mpderivation}-a, the upper frame depicts the plot of dynamic pressure, while the middle frame represents the thermal pressure. The lower frame illustrates the combined total pressure of the solar wind and the dipole pressure. Using spherical coordinates, we derive the shape of the MP at intervals of $20^{\circ}$ along the $\phi$ axis in the equatorial plane as shown in Figure \ref{mpderivation}-b, where the two parts of the MP in Dusk and  Dawn is co-plotted in one frame to account for the asymmetry. The resulting shape exhibits a noticeable asymmetry, with its extension reaching \(5.5\mathrm{R_E}\) towards the Dawn side and \(8.9\mathrm{R_E}\) towards the Dusk side. This asymmetry has been observed and analyzed in several studies. i.e., \citep{Walsh2012,Walsh2014,Haaland2014}. This result helps to explain various magnetospheric phenomena and their effects on space weather, satellite operations, and magnetosphere-ionosphere coupling. }\par
{The position and shape of the magnetopause are not static but vary in response to changes in the solar wind conditions, such as its speed, density, and magnetic field orientation. Monitoring and studying these variations can provide valuable information about the dynamic nature of the Earth's magnetosphere and its response to external disturbances. It can also help improve models and simulations used for space weather forecasting.\citep{Roelof1993}.} \par
\begin{figure}[ht]
	\centering
	\includegraphics[width=0.9\linewidth]{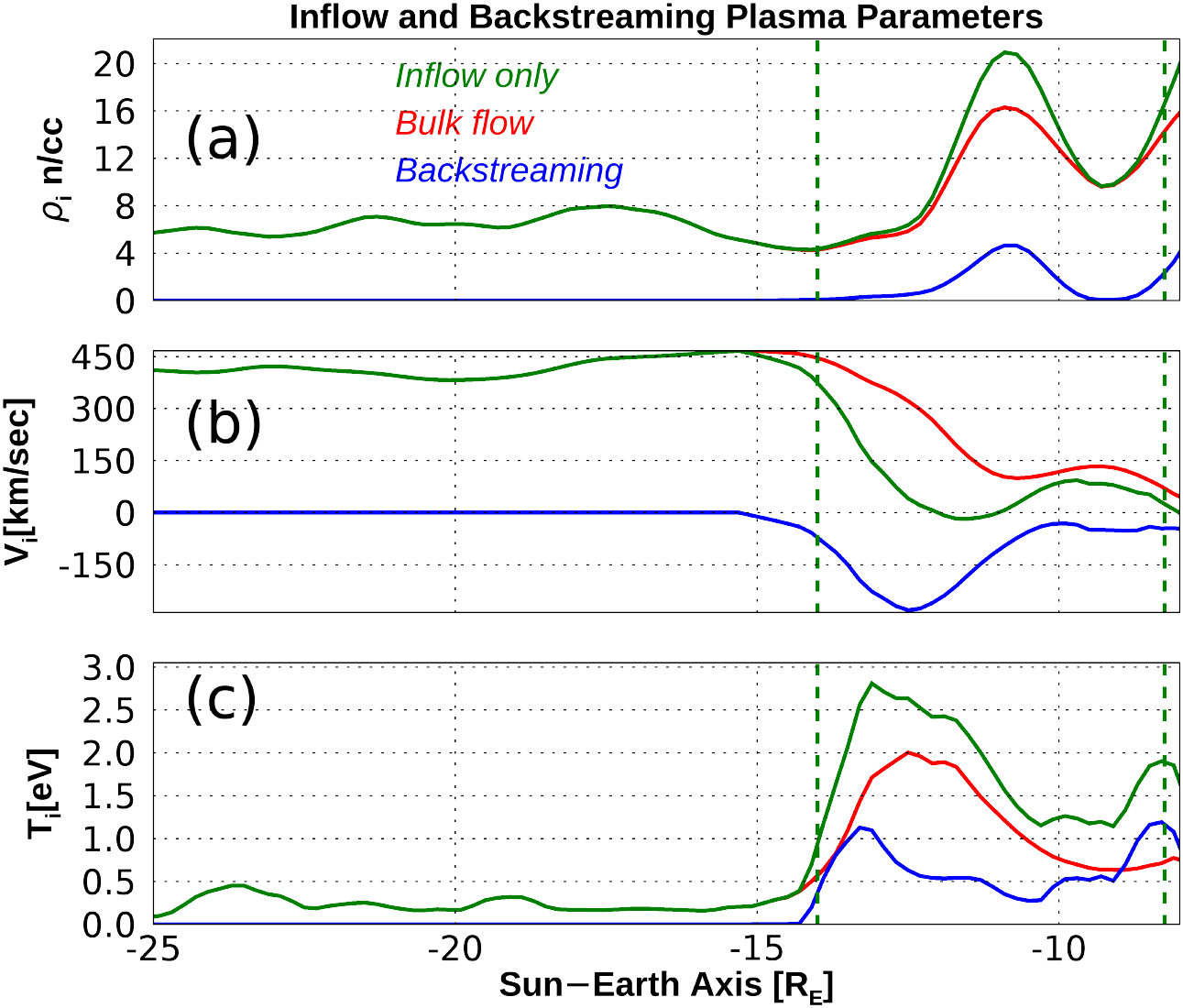}
	\caption{\fontfamily{cmtt} This figure illustrates a comparison between the bulk flow of plasma parameters, the backstreaming ions, and the net inflow of the impinging solar wind particles in terms of their densities, velocities, and temperatures. The data is presented along the Sun-Earth Axis, covering a range from -25 to -8 \(\mathrm{R_E}\). The vertical green lines mark the boundaries of the foreshock and the measured magnetopause  } 
	\label{rvtrev}
\end{figure}
\begin{figure}[ht]
	\centering
	\includegraphics[width=0.9\linewidth]{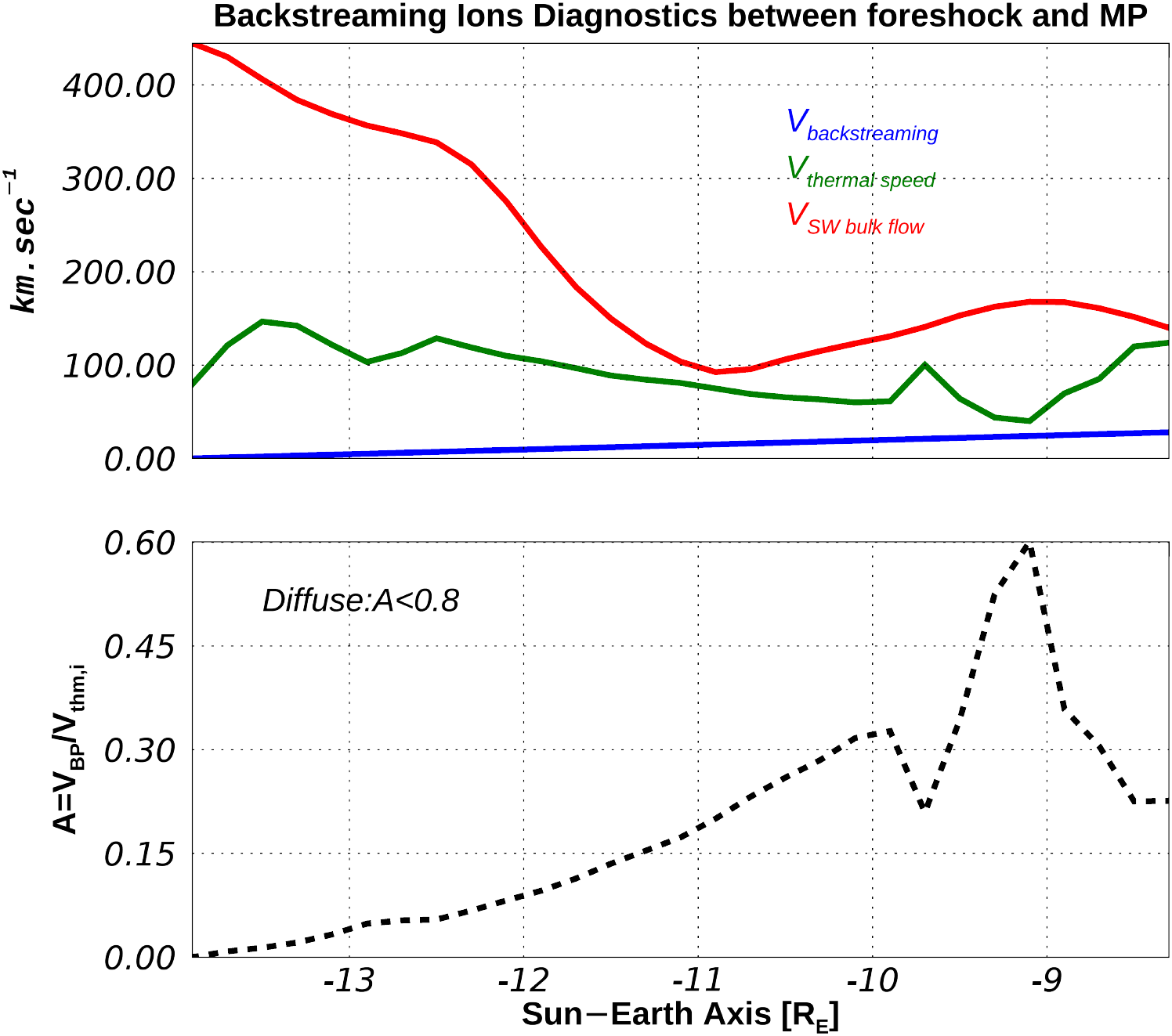}
	\caption{\fontfamily{cmtt} This figure shows the characterization of the backstreaming ions in the region between the MP and the foreshock. In the upper panel, the velocity profile is provided. In the lower panel, the characterization of the backstreaming ions is shown (i.e. \citep{Baraka2021}, \citep{Bonifazi1981}) } 
	\label{fshockcharact}
\end{figure}

  Backstreaming ions upstream of the bow shock is a widely recognized feature in the Earth's magnetosphere, as illustrated in Figure~\ref{mpderivation} \citep{Bonifazi1981,Tanaka1983}.The properties of backstreaming ions for the densities, velocities, and temperatures in the dayside magnetosphere, as presented in Figure \ref{rvtrev}. These ions are a direct consequence of the microscale and macroscale structure of the bow shock. The backstreaming characterization factor reported in \citet{Bonifazi1980,Bonifazi1981,Baraka2021} enabled the analysis of the solar wind backscattered ions at the foreshock region and inside the magnetosheath, as shown in Figures \ref{rvtrev} \& \ref{fshockcharact}, and Table \ref{backvelocity}.\par
Figure \ref{rvtrev}-a shows the backstreaming ions density(blue) appears only at around -13\(\mathrm{R_E}\), which is the location of the foreshock and represents \(\approx 25\%\) of the plasma bulk flow.
The remaining \(\approx 75\%\) bulk flow density (red) shows the foot (foreshock) and the ramp (bow shock) between the foreshock and the MP position. Moreover, eliminating backscattered ions from the bulk flow would increase the dynamic pressure exerted on the MP(green).  Conversely, in Figure \ref{rvtrev}-b, the bulk velocity exhibited deceleration. Eventually, it reached a stagnant state at the MP position at -8.25 \(\mathrm{R_E}\) while the backstreaming velocity reached its maximum value in the foreshock (blue). At the boundaries of the bow shock, the ion temperature experienced an increase(Figure \ref{rvtrev}-c), followed by subsequent cooling down in the magnetosheath.\par
{ The backstreaming ions in the Earth's foreshock can be characterized by analyzing the region delineated by two dashed parallel lines in Figure \ref{rvtrev}, as illustrated (zoomed-in) in Figure \ref{fshockcharact}. The average bulk flow velocity of the solar wind between the foreshock and the MP position is depicted and found to be  \(220 km.sec^{-1}\). %Sona: What does 220 imply? what does it cause?"
The estimated thermal speed of the backstreaming ions is \(\approx 92.4 km.sec^{-1} \) in comparison, their speed is very slow, as seen in Table \ref{backvelocity}.%Sona: What does 220 imply? what does it cause?"
According to the scaling factor proposed by \citep{Bonifazi1980}, all the backstreaming ions present between the foreshock and the MP region can be classified as 100\% diffuse as shown in the lower panel of Figure \ref{fshockcharact}. In other words, the backstreaming ions are characterized by a random distribution, primarily influenced by various factors such as wave-particle interactions. However, no reflected ions were observed in the backstreaming population due to boundaries or discontinuities in the magnetosphere. }\par
\begin{figure}[ht]
	\centering
	\includegraphics[width=0.9\linewidth]{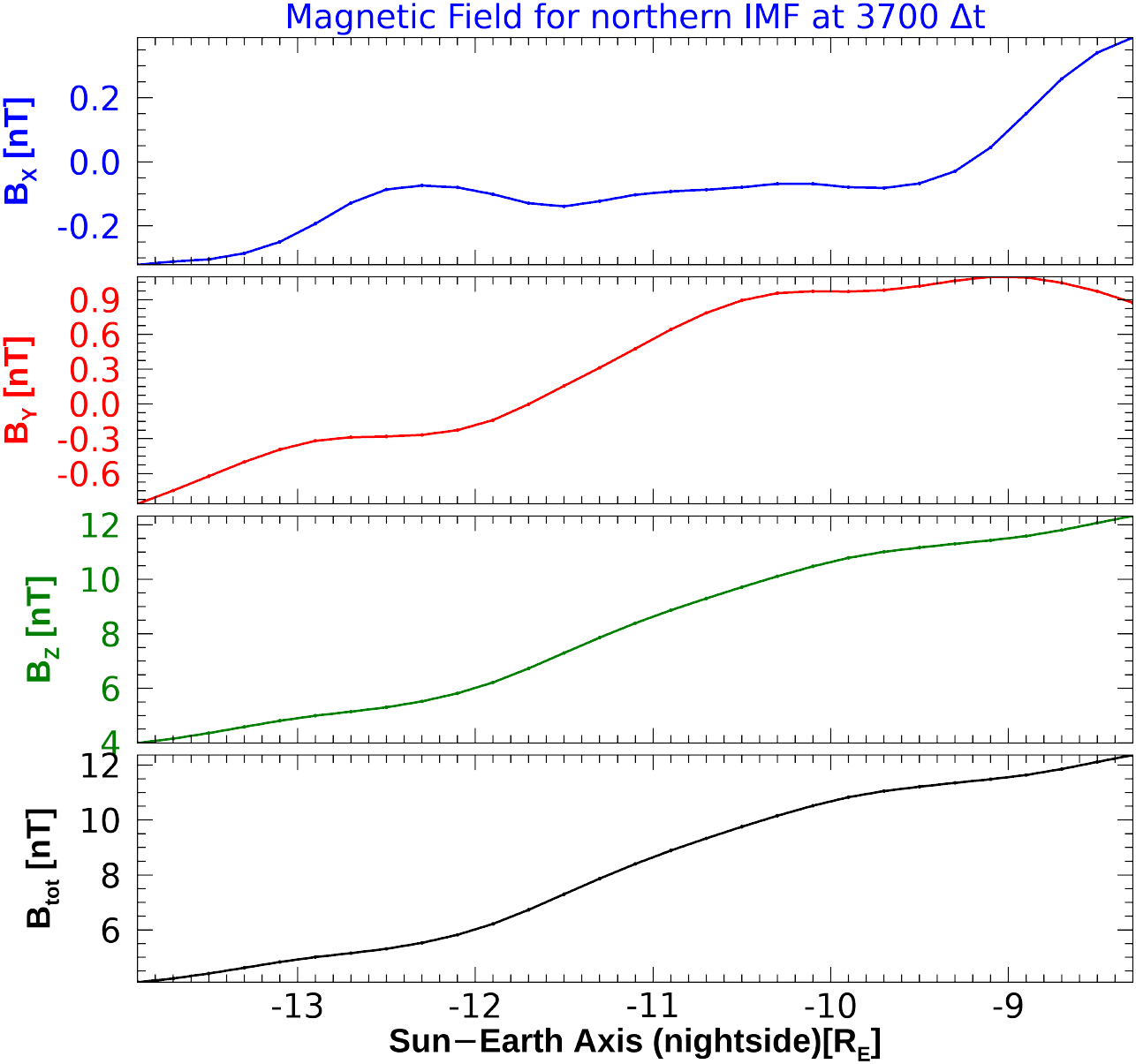}
	\caption{\fontfamily{cmtt} The dayside magnetosphere region depicts the IMF components, indicating that all components, including the total IMF, increase by a factor of 2 in the magnetosheath.  } 
	\label{daysideimf}
\end{figure}

The magnetic field line components within the foreshock region extending up to the MP (magnetopause) position can be seen in Figure \ref{daysideimf}. It shows a change in polarity for  \(\mathrm{B_x}\) and \(\mathrm{B_y}\) from -0.25 to 0.25 nT and -0.7 to 0.9 nT, respectively. Conversely, \(\mathrm{B_z}\) has exhibited an enhancement from 4 to 12 nT, along with an overall increase in \(\mathrm{B_{tot}}\).\par 
The enhanced interplanetary magnetic field (IMF) within the magnetosheath holds critical significance, as it has the potential to signify multiple key phenomena such as magnetic reconnections, alterations in plasma flow patterns, changes in the pressure balance system, and the potential for particle acceleration \citep[e.g.,][]{Vuorinen2019}\par

\begin{figure}[ht]
	\centering
	\includegraphics[width=0.9\linewidth]{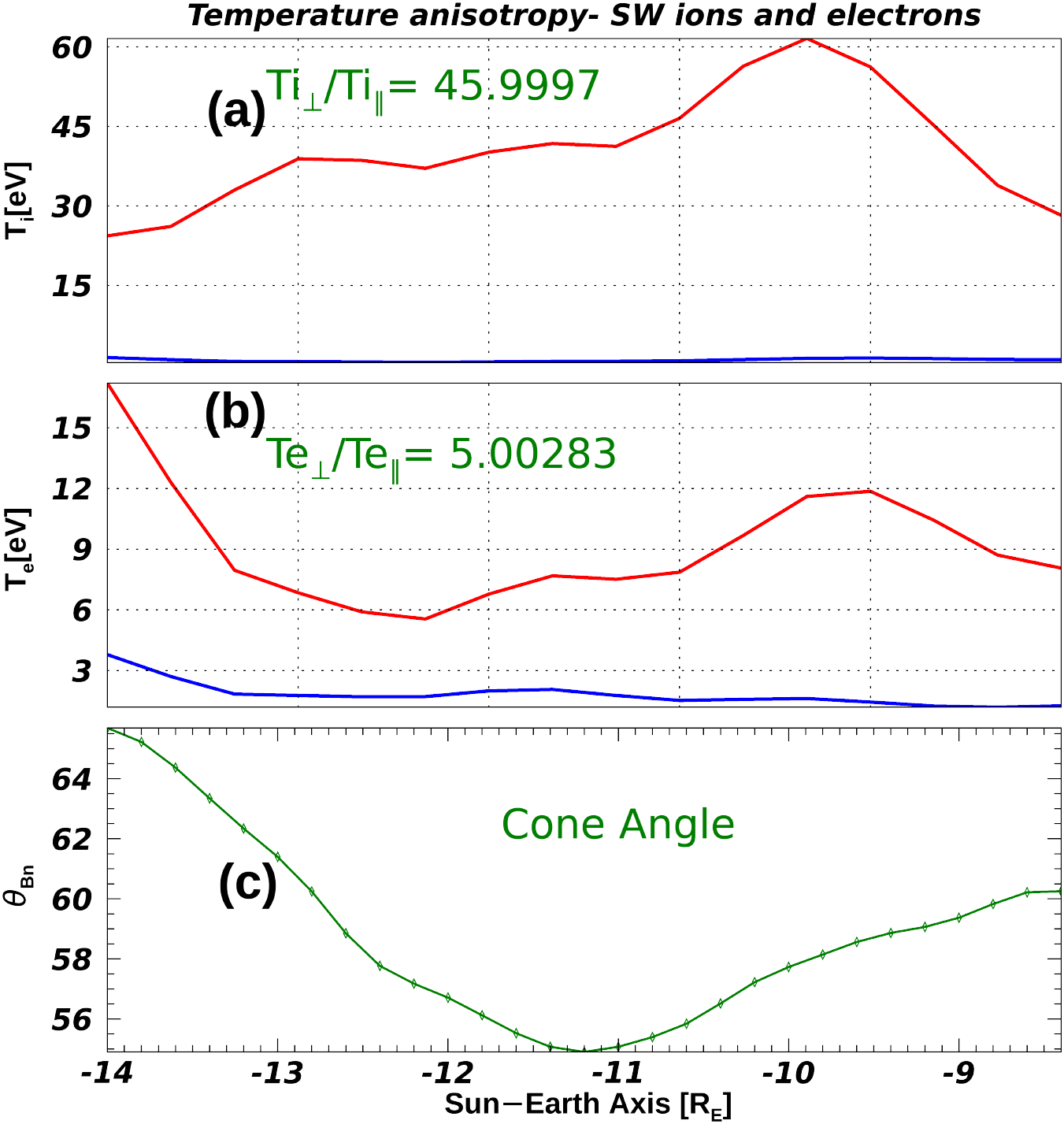}
	\caption{\fontfamily{cmtt} This Figure shows the temperature anisotropy of ions and electrons in the solar wind as observed in the region between the foreshock and the MP position. The parameters are averaged around Y=\(\mathrm{1R_E}\). The lower panel shows the cone angle \(\mathrm{\Theta_{nB}}\), in this case, it is a quasi-perpendicular. It is shown that ions \(\mathrm{T_{\bot}}/T_{\parallel}\) \(\approx 46\), and the ratio for electrons is \(\approx \)5}  
	\label{anisotemp}
\end{figure}

The model incorporates a critical aspect of physical kinetics, specifically the temperature anisotropy of ions and electrons in the solar wind. Figure \ref{anisotemp} provides a visual representation of this temperature anisotropy, with panel (a) depicting the ions' temperature anisotropy, panel (b) representing the electrons' temperature anisotropy, and panel (c) showing the IMF cone angle. The analysis of the data reveals a noticeable contrast in the temperature components. In the case of ions, the perpendicular temperature (\(\mathrm{T_{i\bot}}\) ) is 46 times larger than the parallel components (\(\mathrm{T_{i\parallel}}\) ), as illustrated in Figure \ref{anisotemp}-a. Conversely, for electrons, this ratio is (\(\mathrm{T_{e\bot}}\)/\(\mathrm{T_{e\parallel}}\) = 5 in Figure \ref{anisotemp}-b. The cone angle, as shown in \ref{anisotemp}-c, ranges from 65-55\(\mathrm{^{\circ}}\), indicating a predominantly quasi-perpendicular state throughout the observed range.\par

{To provide a comprehensive analysis of the plasma distribution within the simulation box, we have included Figure \ref{allveldistribution} that depicts the spatial distribution of solar wind ion velocity (represented by the green dotted plot) in 3D. This is overlaid with a linear solar wind plasma velocity that has been averaged by \(1\mathrm{R_E}\) in the Y and Z directions (shown in blue). All values in the figure have been normalized over the initial input value to allow for a clear comparison.}\par

{The spatial distribution of solar wind velocity (\(v_x\)) in the undisturbed solar wind is shown in Figure \ref{allveldistribution}-a. There is no significant change in the direction and/or magnitude of the solar wind velocity until it reaches the foreshock region, where it undergoes deceleration and reversal at a distance of approximately -12\(\mathrm{R_E}\). In contrast, the solar wind velocity (Figure \ref{allveldistribution}-b
\(v_y\)) undergoes significant perturbations and directional changes within the dayside magnetosphere, with only small amounts of reversal observed inside the magnetotail. A similar pattern is observed for (Figure \ref{allveldistribution}-c \(v_z\)). %Sona: What does it mean they are similar? What would have happened if there were differences? "
}\par
\begin{table}[!ht]\caption{At distances encompassing both the foreshock and the MP boundary, the solar wind ion values for inflow, thermal, and bulk backstreaming speeds are averaged. Within this region, the backstreaming ions are entirely diffuse, with no presence of reflected ions or intermediate particles.\citep{Bonifazi1981} }\label{backvelocity} \centering
\vspace*{0.2cm}	\begin{tabular}{|l|c|}
			\hline
			Speed& $km.sec^{-1} $ \\ 
			\hline 		\hline \vspace*{0.1cm}
			$V_{backstreaming}$
			&\textbf{14.0}\\ \hline 	
			\vspace*{0.1cm} $V_{backstreaming_{(thermal)}}$&\textbf{92.4}\\ \hline 	
			\vspace*{0.1cm} $V_{SW_{(bulk)}}$&\textbf{219.4}\\ \hline
	\end{tabular}
\end{table}
\subsection{Escape of The Ionospheric Species Toward the Moon}\label{result2}
Our simulation of magnetosphere-ionosphere coupling(MS-IS) allows us to develop a global picture  of the underlying mechanisms that facilitate energy transfer, particle precipitation, and various phenomena encountered during space weather events. This coupling enables us to track the distribution \(\mathrm{H^+}\) and \(\mathrm{O^+}\)  ions of ionospheric origin throughout the macrostructure of the Earth's magnetosphere. Our current work is aligned with previous research conducted by \citep[e.g.,][]{Harnett2003,Travnivcek2005}. \par

{As explained in \citep{Gronoff2020}, the primary source of \(\mathrm{O^+}\) ions in the magnetosphere of the Earth is the polar outflow, which occurs in specific solar wind conditions and can accelerate the ions above escape velocity. Therefore, a better view of the supply of \(\mathrm{O^+}\) to the Moon will have to consider the conditions where the polar wind is essential and which are outside of the conditions simulated for this paper, which is better viewed as the standard observation in quiet solar activity.} \par 
 {In a study by \citep{Terada2017} observations from the Japanese spacecraft Kaguya allowed to calculate the total number of \(\mathrm{O^+}\) ions on the lunar surface and successfully differentiated their sources, whether it was from magnetospheric or regolith origin. These findings in \citep{Terada2017} suggested the possibility that the Earth’s atmosphere of billions of years ago might have been preserved on the current lunar surface}.\par
\begin{figure}[ht]
	\resizebox{1.\textwidth}{!}{\begin{tabular}{cc}
			
			\includegraphics[width=0.9\linewidth]{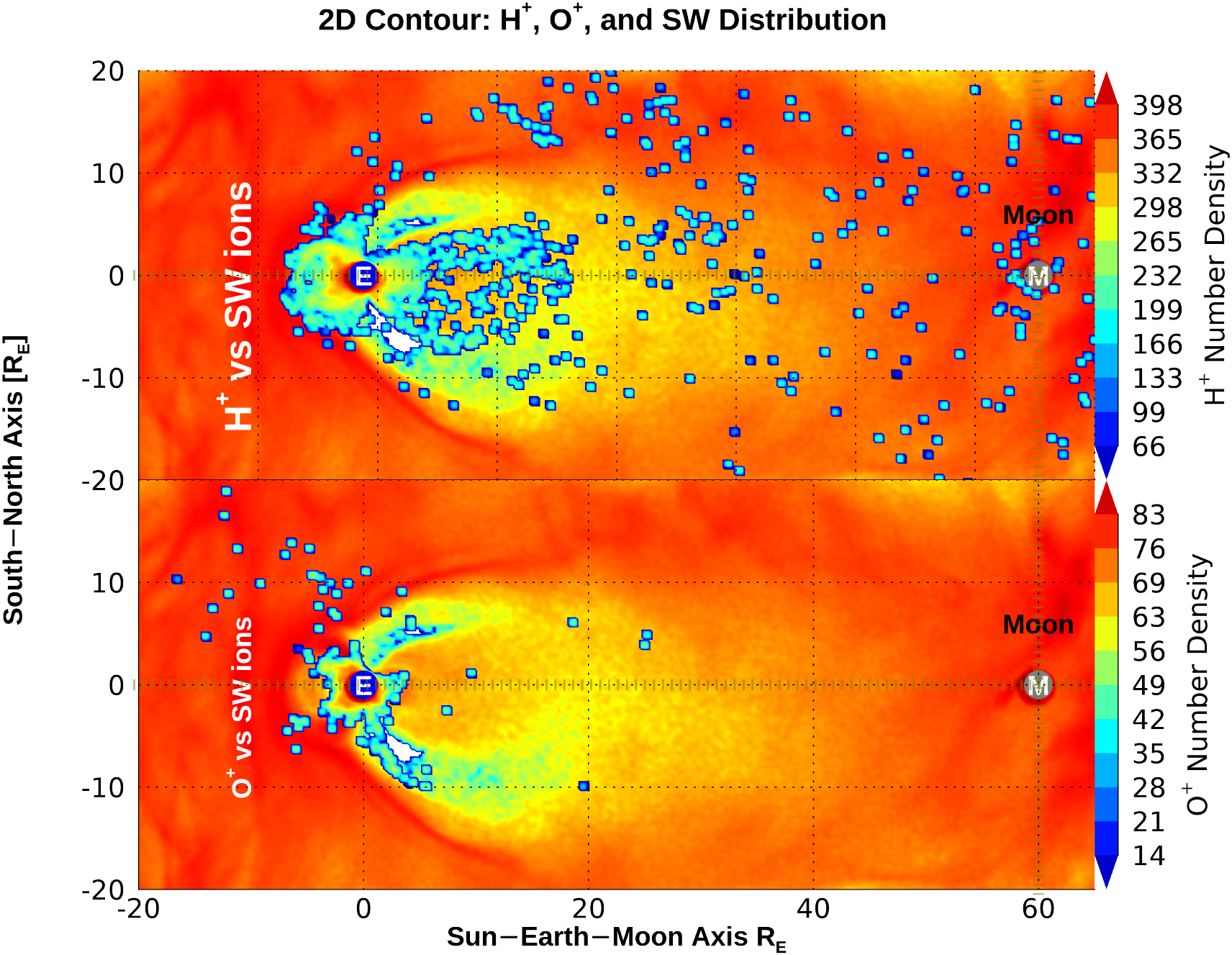}	& \includegraphics[width=0.9\linewidth]{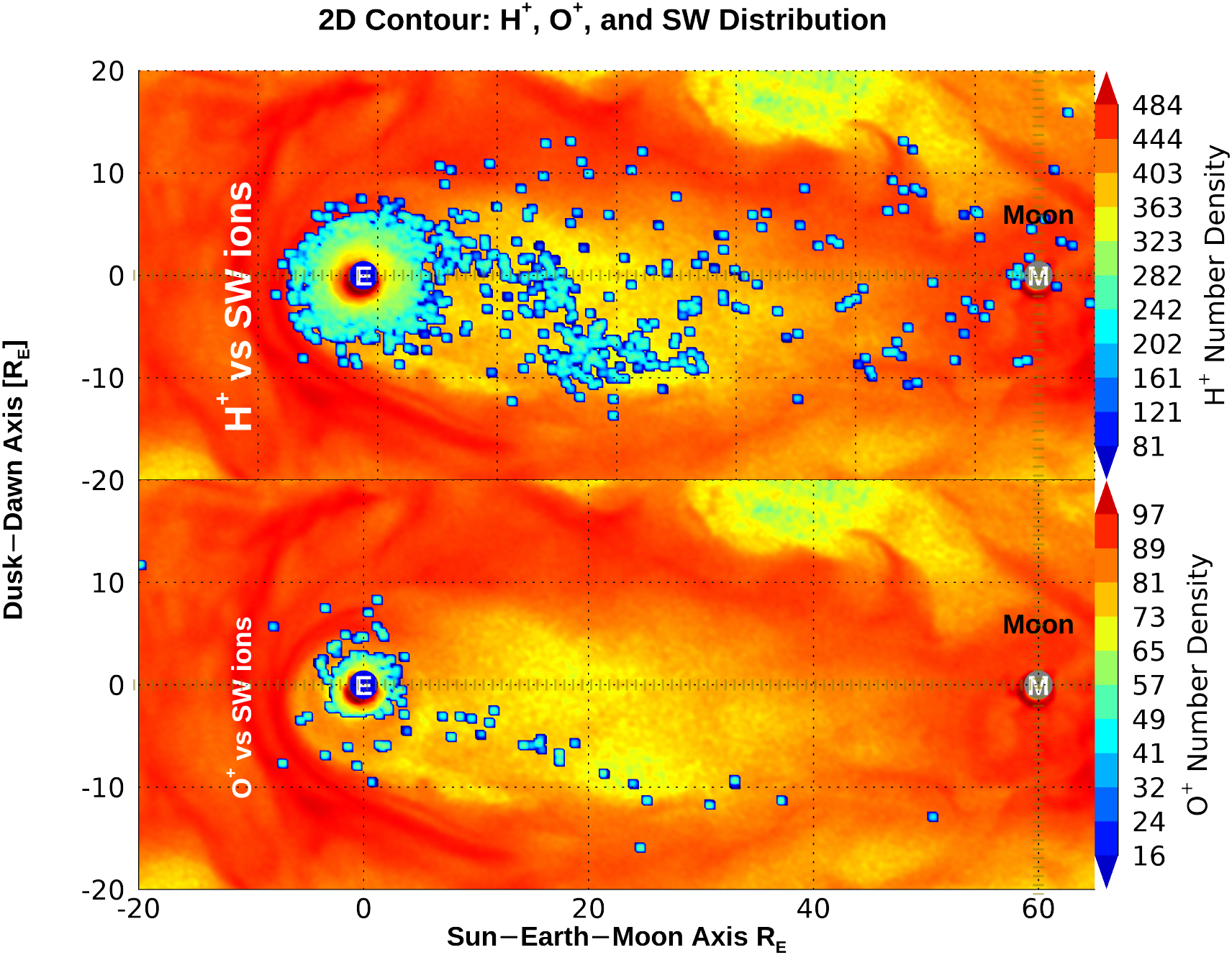}\\
			a&b\\ 
	\end{tabular}}
	
	\caption{{\fontfamily{cmtt}
	Densities of \(\mathrm{H^+}\) (upper frame) and \(\mathrm{O^+}\) (lower frames) of ionospheric origin plotted over a 2D contour of the SW plasma. Panel (a) corresponds to the $\mathrm{XZ-plane}$, panel (b) to the $\mathrm{XY-plane}$.} Density units are 5\(cc^{-1}\) normalized to their initial input value. XZ-contours averaged \(1\Delta=0.2R_E\) in Y,and XY-contours averaged by \(1\Delta=0.2R_E\) in Z )}
	\label{2dallden}
\end{figure}

 
Figure \ref{2dallden} allows visualizing the global distribution of both \(\mathrm{H^+}\) and \(\mathrm{O^+}\) ions over-plotted on the background of the solar wind plasma distribution on two different planes, namely, OX(Noon-Midnight and averaged on Y by 1\(\mathrm{\Delta}\) ) and OY(Equatorial plane: Z by 1\(\mathrm{\Delta}\)). This allows for a comprehensive understanding of the spatial distribution of the  \(\mathrm{H^+}\) and \(\mathrm{O^+}\) ions in both dayside and nightside magnetosphere. \par
\begin{figure}[ht]
	\centering
	\includegraphics[width=0.9\linewidth]{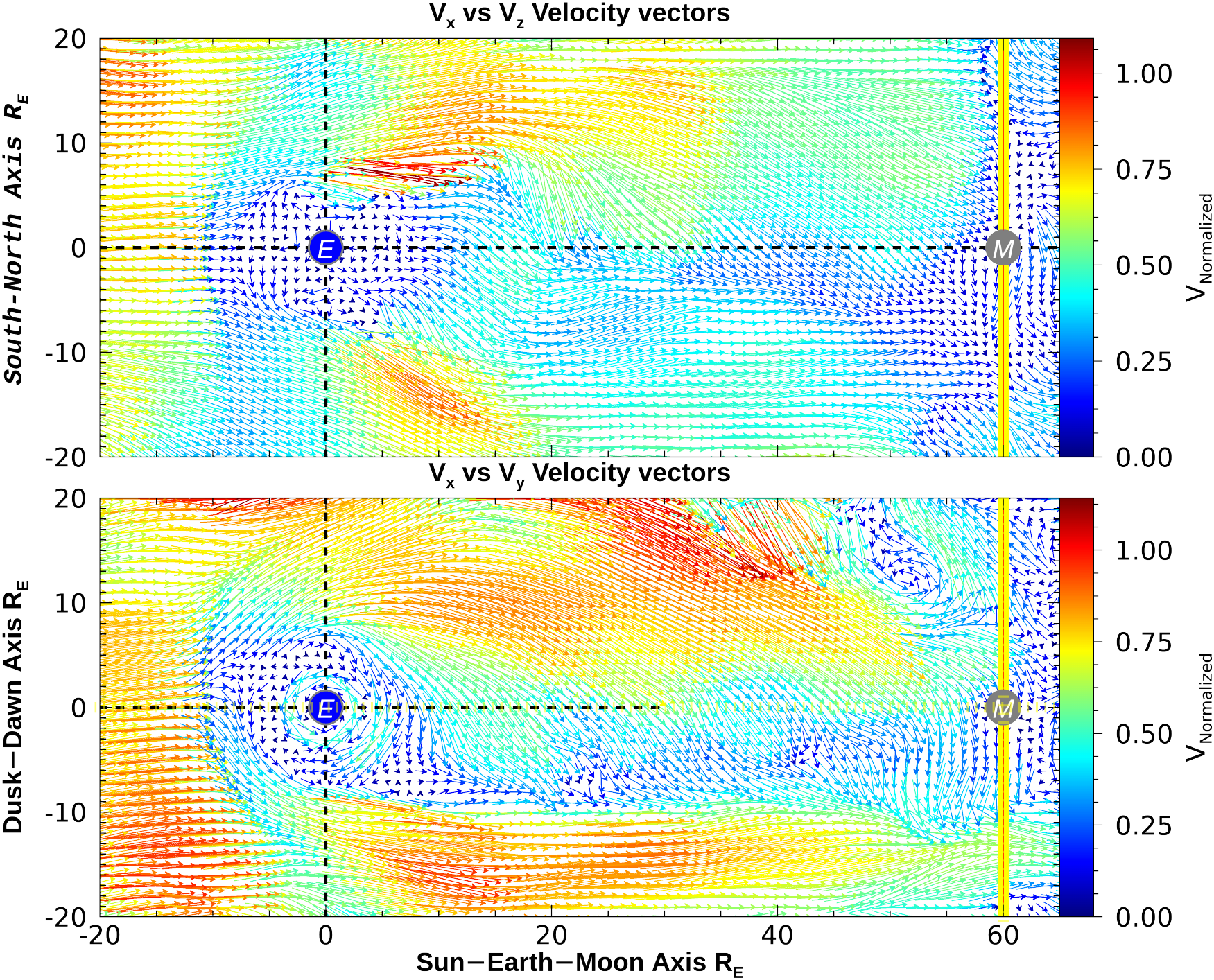}
	\caption{\fontfamily{cmtt} Velocity vector distribution of the solar wind along XZ and XY planes. The arrows indicate the directional flow, and the color code shows their magnitude normalized to the initial solar wind value. Initial IMF input is in the positive Z-direction(Northward IMF)}
	\label{velvector}
\end{figure}
\begin{figure}[ht]
	\centering
	\includegraphics[width=0.9\linewidth]{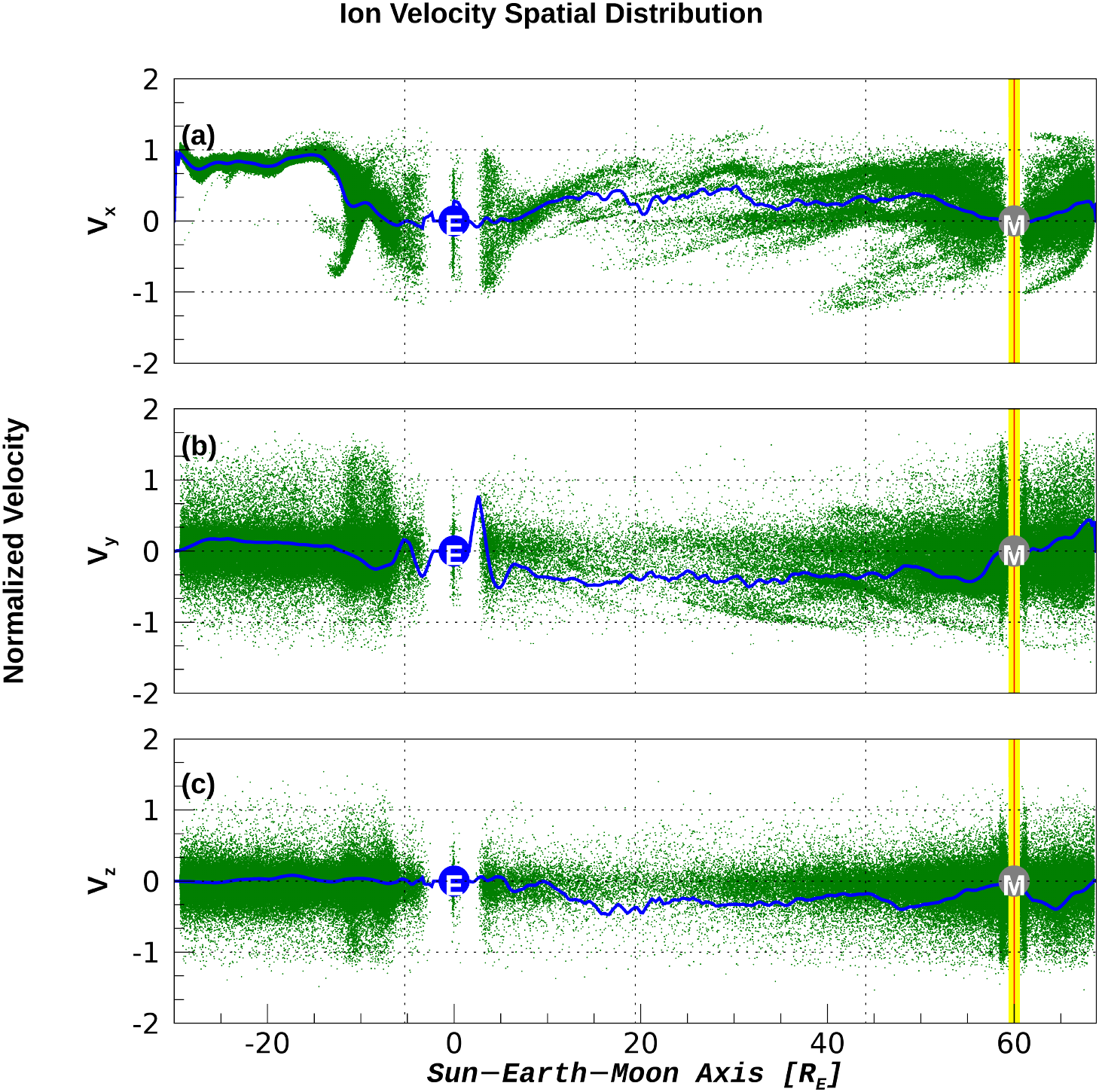}
	\caption{\fontfamily{cmtt} 3-D velocities for the solar-wind ion-test-particle located on the SEM axis (for a simulation box length of $\approx 100R_E$). The averaged solar wind (over $1RE$ in Y and Z) velocity is plotted in blue.}
	\label{allveldistribution}
\end{figure}

Furthermore, Figure \ref{velvector} illustrates the physical flow of plasma in two distinct planes within the simulation box: the noon-midnight plane and the equatorial plane. The velocity vectors, denoted as \(\mathrm{v_x}\)-\(\mathrm{v_z}\)(upper frame) and \(\mathrm{v_x}\)-\(\mathrm{v_y}\)(lower frame), provide valuable insights into the overall plasma dynamics in the large scale magnetosphere. This depiction considers the interconnection between the Sun and Earth, the coupling between the magnetosphere and ionosphere, as well as the interaction between the solar wind and Moon. The significance of this figure lies in its ability to capture large-scale flow patterns and provide a comprehensive understanding of the complex magnetospheric behavior.\par
In \(\mathrm{v_x}\)-\(\mathrm{v_z}\) plane, the velocity vector analysis reveals notable fast flow regions in both northern and southern lobes, spanning 10-25\(\mathrm{R_E}\) on the nightside of the magnetosphere. while on the other hand,  the \(\mathrm{v_x}\)-\(\mathrm{v_y}\) plane depicts that the fast plasma flow is observed on both Dusk and Dawn  flanks. Figure \ref{velvector} provides valuable insights into the plasma dynamics in macrostructure magnetosphere, which gives an instrumental understanding of the magnetosphere's behavior and transport processes. \par
\begin{figure}[ht]\centering
	\resizebox{\linewidth}{!}{\begin{tabular}{cc}
		\includegraphics[width=0.49\linewidth]{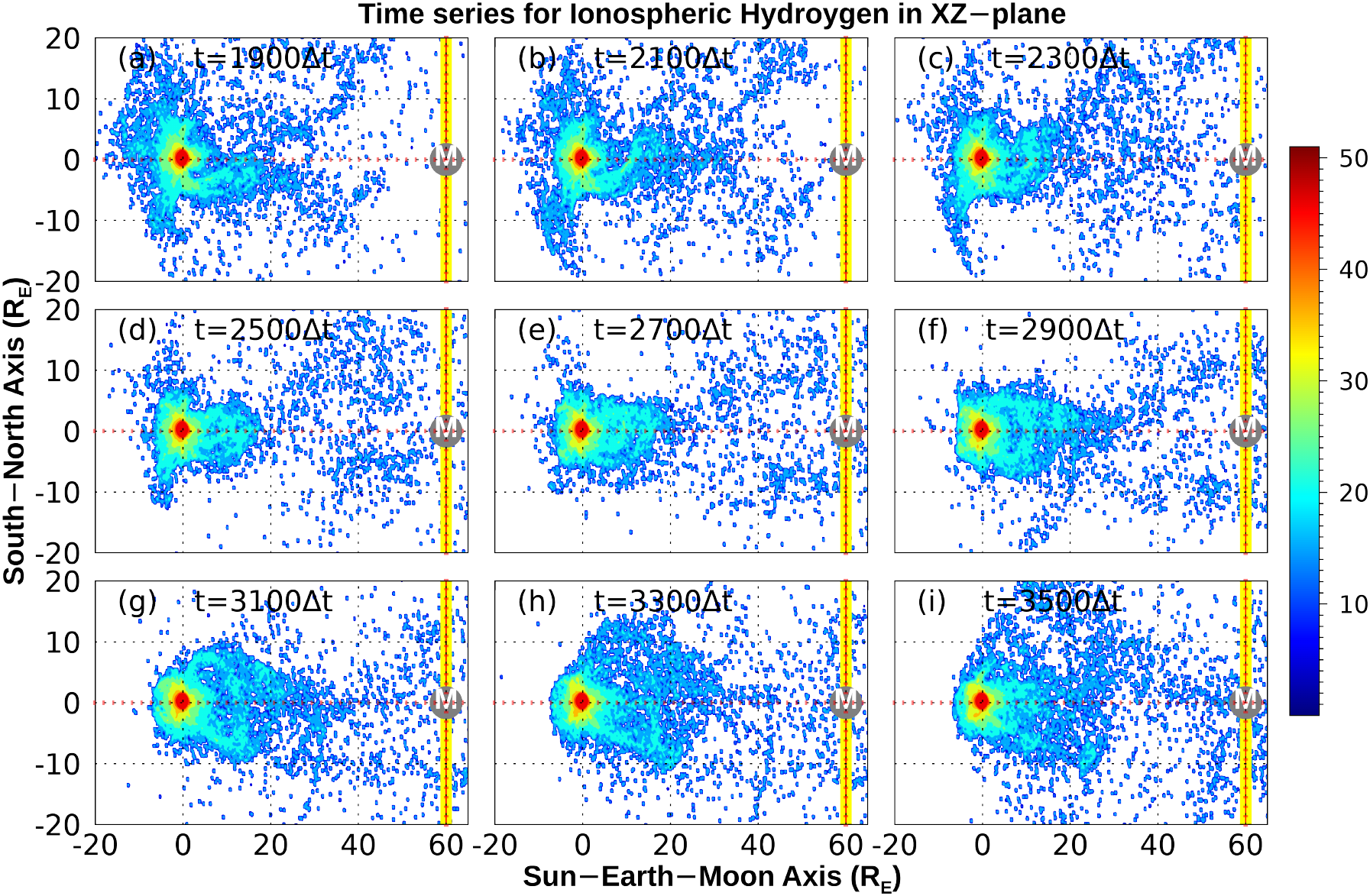}	&\includegraphics[width=0.49\linewidth]{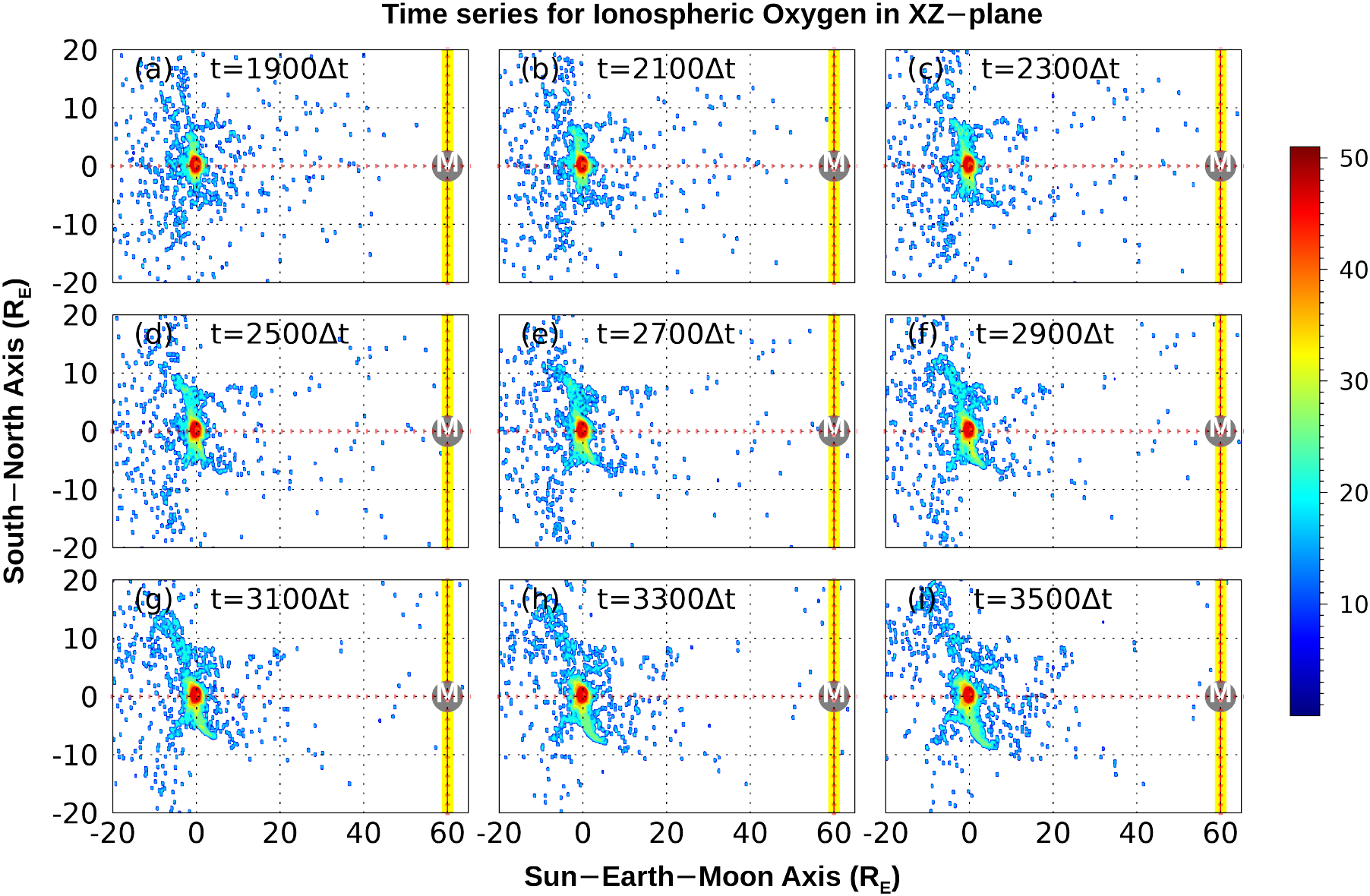}\\
		\includegraphics[width=0.49\linewidth]{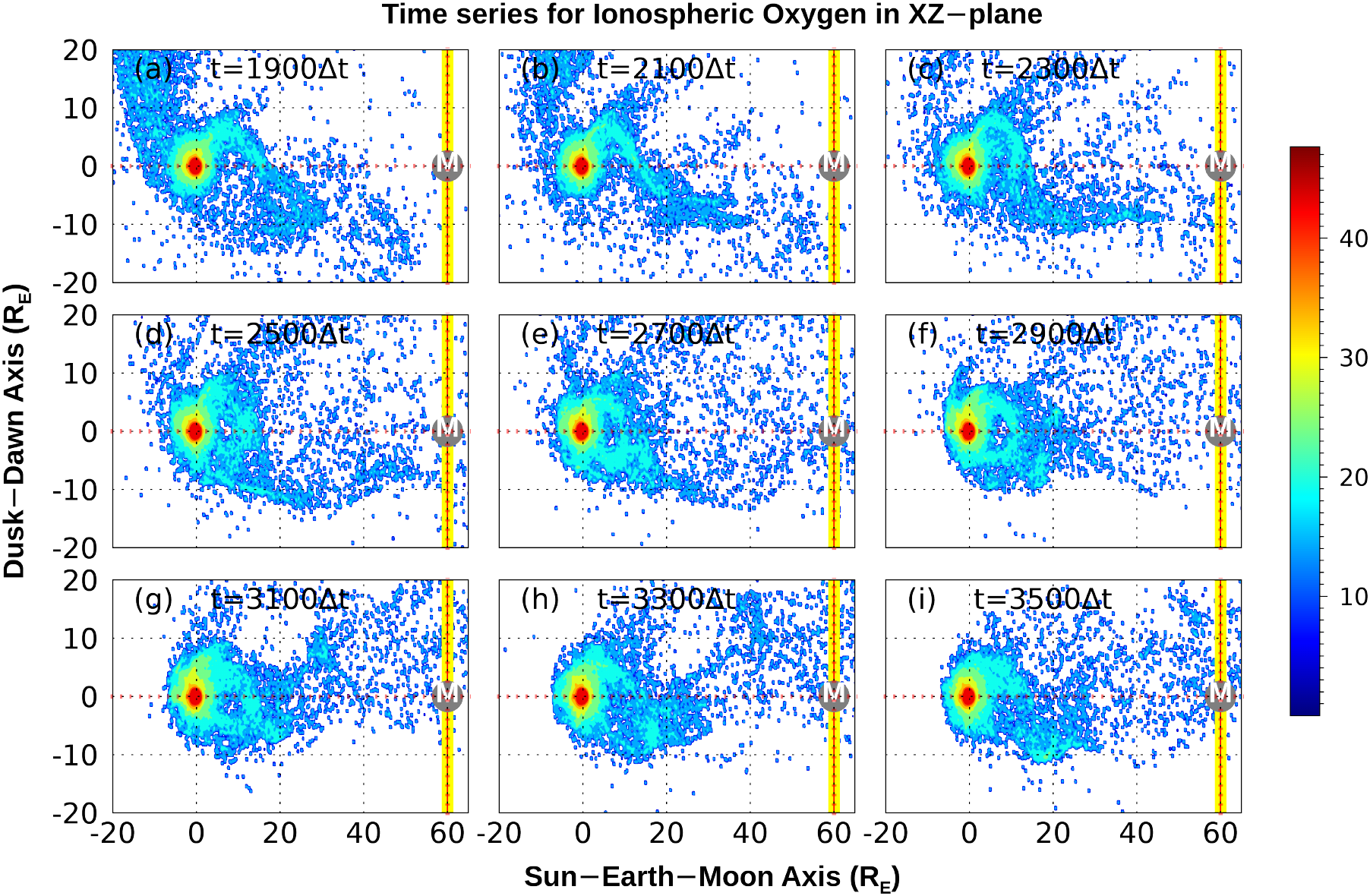}& \includegraphics[width=0.49\linewidth]{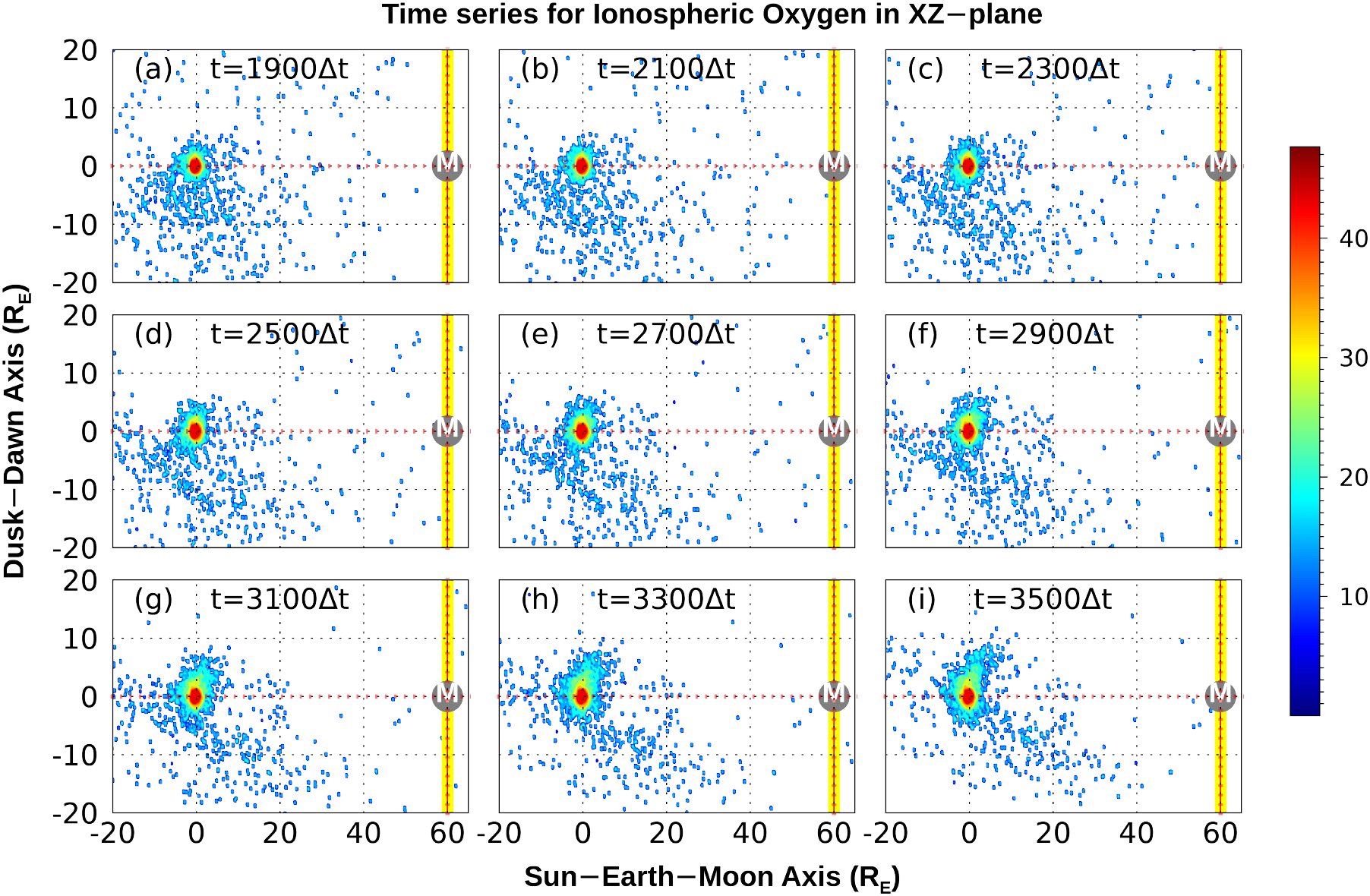} \\
  \includegraphics[width=0.49\linewidth]{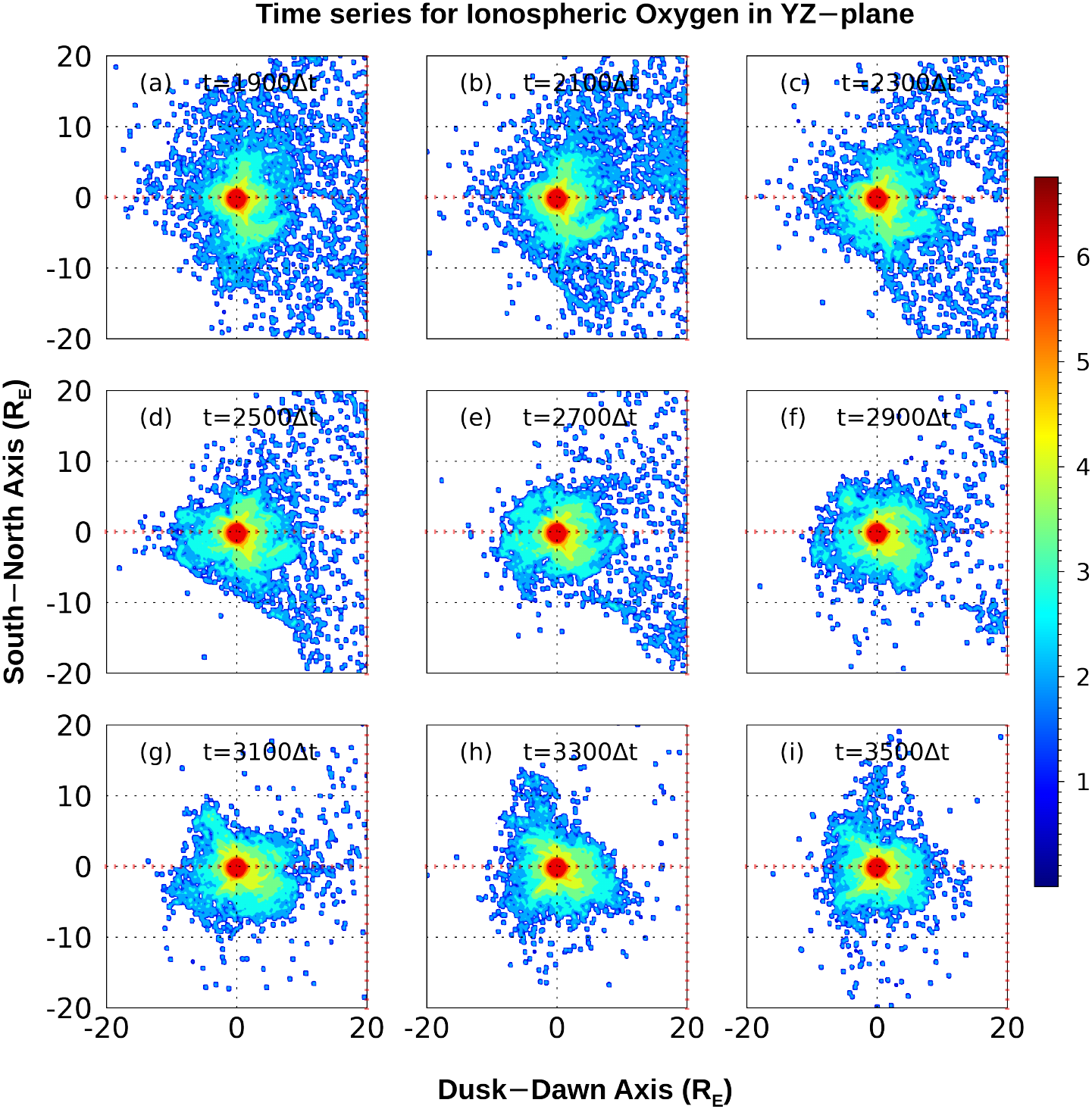}& \includegraphics[width=0.49\linewidth]{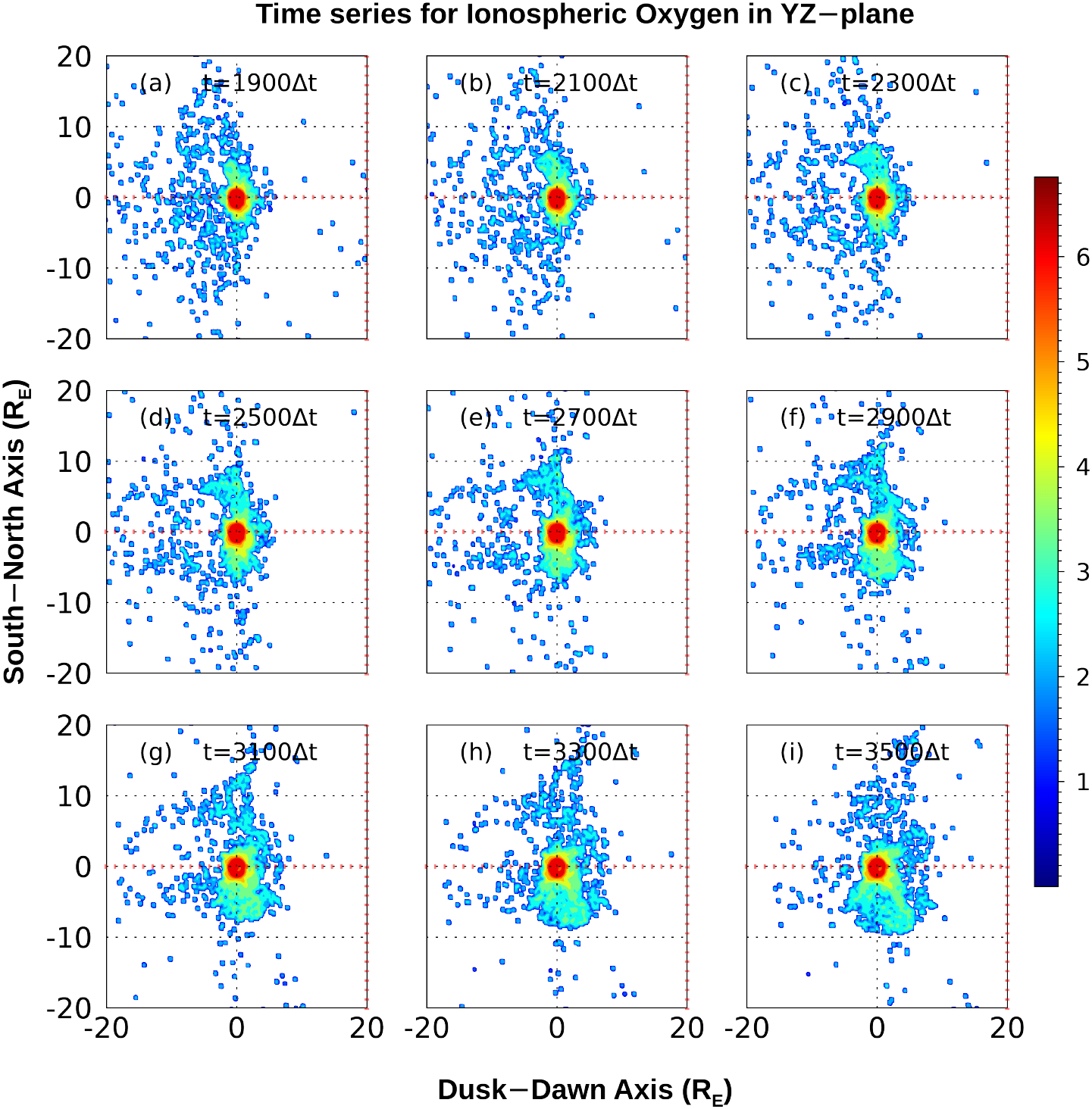}\\
  A&B\\\end{tabular}}  
	\caption{\fontfamily{cmtt}\selectfont This visual representation displays the contours of ionospheric species, namely \(\mathrm{H^+}\) \& \(\mathrm{O^+}\), in a time series format across three planes: XZ, XY, and YZ. The ions are observed escaping towards the magnetosphere from as early as 1900 \(\mathrm{\Delta t}\) to 3500 \(\mathrm{\Delta t}\) at intervals of 100 \(\mathrm{\Delta t}\). The spatial arrangement of the \(\mathrm{H^+}\) \& \(\mathrm{O^+}\) ions suggests that they exit the ionosphere through the polar cap. The figure comprises two panels, A and B, where panel A presents the distribution contours of \(\mathrm{H^+}\) ions in a 3D format, while panel B depicts the same for \(\mathrm{O^+}\) ions.}		
	\label{timeseries}
\end{figure}

To track the escape of \(\mathrm{H^+}\)  and \(\mathrm{O^+}\) ions from the upper ionosphere, we have generated 2D contour plots for both species over a time range spanning from 1900 to 3500 \(\mathrm{\Delta t}\), every 100 \(\mathrm{\Delta t}\), each time step corresponds to 0.64 seconds. Figure \ref{timeseries} shows these contours which are plotted in three distinct planes: Noon-Midnight, Equatorial, and Dusk-Dawn planes. This figure aims to observe the temporal evolution of ion escape patterns within the vast large-scale magnetosphere. These time series contour plots offer valuable information into the dynamics and transport mechanism of these ions as they depart the ionosphere until their final destination. \par  
  { In the upper panel of Figure \ref{timeseries}-a, it is evident that \(\mathrm{H^+}\) ions are escaping the upper ionosphere at t=1900 \(\Delta t\)(which is roughly 20 minutes). This time frame is adequate for particles that do not encounter any interactions with the dipole fields to exit the simulation box from the right simulation boundary. The ions are observed at \(\pm 10\mathrm{R_E}\) along the south-north axis, and approximately \(10\mathrm{R_E}\) towards the dayside. Moreover, there are some infrequent  \(\mathrm{H^+}\) densities observed in the magnetotail, extending up to 40\(\mathrm{R_E}\). } \par
{In contrast, Figure \ref{timeseries}-a right panel shows that \(\mathrm{O^+}\) ions are mostly observed in both the south and north directions, with only a tendency towards the dayside. As time progresses, \(\mathrm{O^+}\)  ions tend to travel towards the dayside, while \(\mathrm{H^+}\)ions escape towards the nightside, reaching the magnetotail and even the Lunar surface, as evident from the XZ-plane time series frames. }\par
{In the XZ plane, the migration of \(\mathrm{H^+}\)ions towards the nightside becomes increasingly apparent over time. Initially, more hydrogen escapes towards the south direction, but the drift is more towards the north by time step 3500 \(\Delta t\). Conversely, as time progresses,\(\mathrm{O^+}\) ions drift towards the dayside with increased diffusion towards the north side. Similarly, the ionospheric behavior of both species is observed in the XY plane, with clear manifestation in the YZ plane. }\par  {This behavior of the ionospheric \(\mathrm{O^+}\) is consistent with  \citet{Barakat2003} in which the impacts of low altitude energization of ions on the dynamics of high latitude plasma was investigated. We see although the escape of light ions \(\mathrm{H^+}\) and \(\mathrm{He^+}\)  are possible when other acceleration mechanisms such as waves and ambipolar fields are small or negligible heavy ions such as \(\mathrm{O^+}\) are bounded by the gravity \citep{Barakat2001}. Several studies cover the \(\mathrm{H^+}\) and \(\mathrm{O^+}\) ions outflow in the polar wind and the aurora  region and their impact on the global dynamics of the magnetosphere macrostructure \citep{Andre1997,Barghouthi1998,Barakat2001,Barghouthi2008,Schunk2009,Nilsson2011,Nilsson2012}.}
\section{Discussion on observational implications}\label{discuss}
Because of the challenges in modeling the whole Earth-Moon system in a single box, the present simulation has to be validated by observations.
On March 8, 2012, \citet{Shang2020} reported a MP tail compression observed by the ARTEMIS spacecraft situated at [x,y,z]=[60.,0.,-5]\(\mathrm{R_E}\). In addition, the solar wind velocity vector downstream did not align with the sun-earth line but had a component along the Y direction. This is in agreement with the solar wind velocity vector modeled here, as seen  in the upper frame of Figure \ref{velvector}. While the compression studied by \citep{Shang2020} is localized in the magnetotail, the present results show that the compression we observed is global in the whole magnetosphere system. Despite this distinction, our findings remain consistent with the reported results in \citep{Shang2020} \par
The moon stays 20\% of each lunar orbit in the magnetotail \citep[e.g.,][]{Terada2017}), enabling the examination of atmospheric escape ions (specifically the upper ionosphere) originating from Earth \citep{Dandouras2023}.
Considering the initial conditions established for the ionospheric species, as discussed in \citep{Bilitza2008,Bilitza2011}, as well as the results of the present simulations, it appears that an insufficient amount of oxygen is observed to reach the surface of the moon. The ionospheric species lack the necessary upward velocity to overcome the gravitational pull and the presence of closed magnetic field lines. However, noteworthy observations from the Japanese spacecraft Kaguya, as reported by \citep{Terada2017}, indicate a significant presence of \(\mathrm{O^+}\)  ions. Additionally, in Table 1 of \citet{Terada2017} the net flux of \(\mathrm{O^+}\) ions observed during various time intervals was quantified. Therefore, the existence of \(\mathrm{O^+}\) ions from Earth reaching the lunar surface shows that some of the assumptions used in the present modeling are too simplistic; the most notable approximation being the null vertical velocity of O$^+$ from the polar wind.\par 
\begin{table}[!ht]
\resizebox{\textwidth}{!}{	\begin{tabular}{caayssy}
\rowcolor{yellow}& \multicolumn{2}{c}{\underline{Oxygen}}&&\multicolumn{2}{c}{\underline{Hydrogen}}&\\
\rowcolor[gray]{0.85}&Day&Night&&Day&Night&\\
\rowcolor[gray]{0.85}	Time step	\(\mathrm{\Delta t}\)&A&B&A/B&C&D&C/D\\ \hline\hline
        1900&       2.75\(\times 10^4\)&       1.14\(\times 10^4\)&           2.40&      2.36\(\times 10^5\)&      1.14\(\times 10^5\)&           2.07\\
	2100&       2.89\(\times 10^4\)&       1.18\(\times 10^4\)&           2.45&      2.44\(\times 10^5\)&      1.15\(\times 10^5\)&           2.12\\
	2300&       2.99\(\times 10^4\)&       1.26\(\times 10^4\)&           2.37&      2.49\(\times 10^5\)&      1.25\(\times 10^5\)&           1.99\\
	2500&       3.06\(\times 10^4\)&       1.35\(\times 10^4\)&           2.27&      2.53\(\times 10^5\)&      1.33\(\times 10^5\)&           1.90\\
	2700&       3.15\(\times 10^4\)&       1.5\(\times 10^4\)&             2.10&      2.57\(\times 10^5\)&      1.42\(\times 10^5\)&           1.80\\
	2900&       3.19\(\times 10^4\)&       1.55\(\times 10^4\)&           2.05&      2.60\(\times 10^5\)&      1.41\(\times 10^5\)&           1.83\\
	3100&       3.43\(\times 10^4\)&       1.68\(\times 10^4\)&           2.03&      2.52\(\times 10^5\)&      1.52\(\times 10^5\)&           1.66\\
	3300&       3.67\(\times 10^4\)&       1.78\(\times 10^4\)&           2.06&      2.65\(\times 10^5\)&      1.55\(\times 10^5\)&           1.70\\
	3500&       3.83\(\times 10^4\)&       1.87\(\times 10^4\)&           2.04&      2.76\(\times 10^5\)&      1.59\(\times 10^5\)&           1.73\\ 
3700& 4.05\(\times 10^4\)&2.02\(\times 10^4\)&2.00&2.75\(\times 10^5\)&1.73\(\times 10^5\)&1.58\\        
 \hline \hline
	\end{tabular}}\vspace{0.2cm}
	\caption{The outflow rates of both \(\mathrm{H^+}\) and \(\mathrm{O^+}\) are continuously increasing over time on both the day and night sides of the planet. As a function of time steps, measured at intervals of 200 time steps (\(\mathrm{\Delta t}\)) the ratio between the two species shows a tendency to decrease towards a stable state. The escape rate of ionospheric species exhibits a monotonic behavior and eventually reaches a stable state. Figure \ref{timeseries} illustrates the observed trend, providing visual evidence of the progressive increase in the outflow rates of both \(\mathrm{H^+}\) and \(\mathrm{O^+}\) over time on both the day and night sides of the planet. 
}\label{metricratio}
\end{table}
The PIC code boasts a distinctive feature of incorporating gravitational force. The migration behavior of \(\mathrm{O^+}\) ions towards the dayside magnetosphere and \(\mathrm{H^+}\) ions towards the magnetotail can be seen in Figure~\ref{timeseries}. Remarkably, both \(\mathrm{H^+}\) and \(\mathrm{O^+}\) ions outnumber their respective counterparts on the nightside. Table \ref{metricratio} provides a comprehensive overview of the total number of \(\mathrm{O^+}\) ions on both day and nightside, highlighting their corresponding ratios (highlighted in yellow). It is noteworthy that the number of \(\mathrm{O^+}\) ions increases steadily over time, a decreasing ratio appears between the day and night sides and tends towards a stable ratio. This implies a stable escape rate for \(\mathrm{O^+}\) ions and means that the code evolved towards a steady state, allowing us to discuss the day/night asymmetries. Conversely, the total number of  \(\mathrm{H^+}\) ions on the dayside consistently surpasses that on the nightside. However, owing to their distribution along the open magnetic field lines in the magnetotail, \(\mathrm{H^+}\) ions manage to overcome gravity and are drawn closer to the lunar surface. \par

A major question is how these results compare with the observations made by the Magnetospheric Multiscale (MMS) mission, as reported in \citet{Zeng2020}. It is worth noting that our code utilizes coordinates as described and depicted in \citep[e.g., Figure 1][]{Baraka2021}; the positive direction corresponds to dawn, while the negative direction corresponds to dusk. On the other hand, the {GSM}(Geocentric Solar Magnetospheric) coordinates used in \citep{Zeng2020} are exactly the opposite. The study by \citep{Zeng2020} demonstrates that the density of  \(\mathrm{O^+}\)  ions tends to be more drifted towards the dusk side during periods of southern interplanetary magnetic field (IMF) with positive IMF \(\mathrm{B_y}\) orientation. This same asymmetry trend appears in the present simulations (see Figure \ref{timeseries}) during periods of northward IMF with negative IMF \(\mathrm{B_y}\) orientation. This consistent agreement between our PIC code and the MMS data provides a solid foundation for future work incorporating ARTEMIS and {MMS} data to continue studying the Sun-Earth-Moon Coupling.\par
The computation allowed the assessment of \(\mathrm{O^+}\) ion accelerations in the tail plasma sheet under the influence of a northern interplanetary magnetic field (IMF) and strong solar wind dynamic pressure. Those compare well with the observations reported in the statistical study conducted by \citep{Kronberg2015} based on Cluster spacecraft data. One of the noteworthy limitations in our study pertains to the relatively low input ionospheric numbers of oxygen and hydrogen ions. The energization levels observed were insufficient to counteract gravity and enable escape from bounded closed magnetic field lines. The O$^+$ vertical velocity at the top of the ionosphere was supposed to be negligible, which is not what atmospheric escape models are showing. In polar regions, O$^+$ can be accelerated above escape velocity and therefore may more easily reach the Lunar orbit along the magnetotail.
These factors demand ongoing attention in future simulations to enhance our comprehension of this phenomenon and are a limiting factor that needs to be strongly highlighted.\par
\clearpage
\section{Conclusion and Future Work}\label{conclusion}
The first part of the present global 3D kinetic simulation focuses on the coupling between the Sun, magnetosphere, ionosphere, and Moon. Specifically, the analysis is focused on the Sun-Earth connection on the dayside and the coupling between the large-scale magnetosphere and ionosphere. A significant result  is the determination of the size and shape of the magnetopause (MP) (see also \citet{Baraka2021}) to comprehend how mass, momentum, and energy are transferred from the solar wind to the inner magnetosphere, as well as in the field of space weathering. The analysis revealed that the MP asymmetrically extends to 8.25\(\mathrm{R_E}\) sunward, 5.5\(\mathrm{R_E}\) Dawn-ward and 8.9\(\mathrm{R_E}\) Dusk-ward in spherical polar coordinates on the XY plane. These results show the MP's shape along the equatorial plane varied significantly, with a clear asymmetry between the Dusk and Dawn directions, as we derived its shape every 20 \(^{\circ}\).\par
 Another notable achievement is the successful coupling between the IAPIC code \citep{Baraka2021,Ben-Jaffel2021} and the International Reference Ionosphere (IRI) code \citep{Bilitza2008,Billingham2011}. This coupling allowed  to track the distribution of the escaped \(\mathrm{H^+}\) and \(\mathrm{O^+}\) ions from the upper ionosphere through the complex structure of the magnetotail up to the surface of the Moon (as described in Paper II). These findings indicate that \(\mathrm{H^+}\) ions from the ionosphere tend to migrate towards the night side (in the Lunar direction), while \(\mathrm{O^+}\) ions escape towards the dayside through the polar caps.\par
  By conducting PIC simulations of escaped oxygen ions from the upper ionosphere, we have validated the distribution pattern of oxygen near the Earth as reported in the study by \citep{Kronberg2015}. The agreement between our PIC simulations and the observations is clearly depicted in Figure \ref{timeseries}. Both our PIC simulations and the statistical analysis performed in \citep{Kronberg2015} reveal a substantial increase in oxygen intensity towards the dusk side of the near Earth region. This trend is observed during periods of supersonic implied dynamic pressure in our case and dynamic pressure variations in the study by \citep{Kronberg2015}.\par
  
One of the key findings of our study pertains to the spatial asymmetrical distribution of \(\mathrm{O^+}\) ions escaping from the upper ionosphere, which predominantly occurs towards the dusk side. This finding is in line with the observations reported by \citep{Zeng2020} based on data obtained from the Magnetospheric Multiscale (MMS) mission. This significant result contributes to advancing our comprehension of the Magnetosphere-Ionosphere-Moon coupling. By leveraging this understanding, we can better analyze and interpret events captured by MMS and ARTEMIS data, thereby deepening our knowledge of the underlying physics involved in this crucial coupling.  \par 
Other findings of this study show how and where the solar wind in the dayside magnetosphere has developed anisotropic behavior for both ions and electrons temperature, which concurrently occurs with the characterization of the backstreaming ions. The backstreaming ions' velocities and properties are shown in Table \ref{backvelocity} 
 {This research has opened up a vast horizon for us to continue using the developed techniques in various fields, including Earth's magnetospheric physics \citep{Baraka2021,Ben-Jaffel2021} and has a potential impact on exoplanet studies \citep{Gronoff2020,Ben-Jaffel2021}.
The current results show that it may be important to further investigate the dynamics of the cusp in the dayside magnetosphere and study the complex current systems in the magnetotail. In addition, it highlights our plan to investigate further the coupling between the magnetosphere and ionosphere to track oxygen and hydrogen ions with varying energy levels, allowing a better simulation of the abundance of these ions at the lunar surface. %We also intend to utilize ionization codes to calculate water molecules' potential formation from these earth-origin particles at the lunar surface.
 These  findings align our model with prominent models used for simulating and regenerating data from various space missions to the Moon, allowing further lunar studies. One example of such studies can be found in Paper II, where the impact of the magnetotail on the H$^+$ density is studied in detail along with surface charging.}
\clearpage
\singlespacing
\bibliography{biblio.bib}\label{ref}
\clearpage
\section{Acknowledgment}\label{ack}
This research was carried out at the National Institute of Aerospace, NASA Langley Research Center, Science Systems and Applications Inc., and partially at the Jet Propulsion Laboratory, California Institute of Technology, under a contract with the National Aeronautics and Space Administration Contract Number 80NM0018D0004. We express our gratitude for the valuable code contribution provided by Lotfi ben Jaffel. We gratefully acknowledge the assistance provided by Bjorn Davidsson, Larry Paxton, Richard Barkus, and Enrico Piazza for their immeasurable support and valuable discussions on the issues and progress of this research. The work of G.G. was financed by NASA contract 80NM0018D0004 and SSAI IRAD funding.
\clearpage  %Lire proposal 
\section{Figures}\label{figures}
%%% Figure 1

% Figure 2

%% Figure3===
%% Figure4===

%%% Figure 5

%% Figure6===

%%%%   Figure 7 ===========

%%%%%%Fig 8
\centering

%%%%   Figure 9===========

\end{document}